\documentclass[conference]{IEEEtran}
\IEEEoverridecommandlockouts
\usepackage{cite}
\usepackage{amsmath,amssymb,amsfonts}
\usepackage{graphicx}
\usepackage{textcomp}
\usepackage{xcolor}
\usepackage[hyphens]{url}
\usepackage{soul}
\usepackage{color}
\usepackage[]{siunitx}
\usepackage{subfig}
\usepackage{multirow}
\usepackage[normalem]{ulem}
\usepackage{amsmath}
\usepackage{algorithmicx}
\usepackage{algorithm} 
\usepackage{algpseudocode} 
\usepackage{xspace}
\usepackage{xurl}
\usepackage[font=small,labelfont=bf]{caption}

\usepackage{hyperref}

\newcommand{\fig}[1]{Figure~\ref{#1}}

\newcommand{\sect}[1]{Section~\ref{#1}}
\newcommand{\tab}[1]{Table~\ref{#1}}
\newcommand{\algo}[1]{Algorithm~\ref{#1}}
\newcommand{\proposed}[0]{ElasticRec\xspace}

\algnewcommand{\LineComment}[1]{\State \(\triangleright\) #1}

\algblock{ParallelFor}{EndParallelFor}
\algnewcommand\algorithmicparfor{\textbf{parallel for}}
\algnewcommand\algorithmicpardo{\textbf{do}}
\algnewcommand\algorithmicendparfor{\textbf{end\ parallel for}}
\algrenewtext{ParallelFor}[1]{\algorithmicparfor\ #1\ \algorithmicpardo}
\algrenewtext{EndParallelFor}{\algorithmicendparfor}

\def\BibTeX{{\rm B\kern-.05em{\sc i\kern-.025em b}\kern-.08em
    T\kern-.1667em\lower.7ex\hbox{E}\kern-.125emX}}
\begin{document}

\title{\LARGE ElasticRec: A Microservice-based Model Serving Architecture\\ Enabling Elastic Resource Scaling for Recommendation Models
\thanks{This is the author preprint version of the work. The authoritative version will appear in the Proceedings of the 51st IEEE/ACM International Symposium on Computer Architecture (ISCA-51), 2024.}
}

\author{
\IEEEauthorblockN{Yujeong Choi}
\IEEEauthorblockA{School of Electrical Engineering\\
KAIST\\
\textit{yjchoi0606@kaist.ac.kr}}
\and
\IEEEauthorblockN{Jiin Kim}
\IEEEauthorblockA{School of Electrical Engineering\\
KAIST\\
\textit{jiin.kim@kaist.ac.kr}}
\and
\IEEEauthorblockN{Minsoo Rhu}
\IEEEauthorblockA{School of Electrical Engineering\\
KAIST\\
\textit{mrhu@kaist.ac.kr}}
}

\maketitle


\begin{abstract}
With the increasing popularity of  recommendation systems (RecSys), the demand
for compute resources in datacenters has surged. However, the model-wise
resource allocation employed in current RecSys model serving architectures
falls short in effectively utilizing resources, leading to sub-optimal total
cost of ownership. We propose \proposed, a model serving
architecture for RecSys providing resource elasticity and high memory
efficiency. \proposed is based on a microservice-based software architecture
for fine-grained resource allocation, tailored to the heterogeneous resource
demands of RecSys. Additionally, \proposed achieves high memory efficiency via
our utility-based resource allocation. Overall, \proposed achieves an
average 3.3$\times$ reduction in memory allocation size and 8.1$\times$
increase in memory utility, resulting in an average 1.6$\times$ reduction in
	deployment cost compared to state-of-the-art RecSys inference serving
	system.

\end{abstract}


\section{Introduction}

Deep neural network (DNN) based recommendation system (RecSys) accounts for a significant portion of
machine learning (ML) inference cycles in modern datacenters (75\% in
		Meta~\cite{gupta2020architectural}, 25\% in Google~\cite{tpu4}).  
These latency-critical
ML services operate with stringent service level agreement (SLA) goals on tail
latency, so maximizing latency-bounded throughput (i.e., number of service
		queries processed per second that meets SLA, aka QPS) becomes critical.

To serve billions of service queries around the world, datacenters replicate a
large fleet of inference servers, each server replica provisioned with
its own copy of the entire model
parameters~\cite{deeprecsys,hercules,merlin,gupta2020architectural}. Such
baseline ``model-wise'' resource allocation enables each server replica to independently service user queries, which helps utilize
query-level parallelism across the fleet and improve QPS.  However, a critical
limitation of model-wise resource allocation is that the way resources are
allocated does not consider how well it is actually utilized, leading to
significant waste in resources.  This work identifies two key
reasons behind the baseline's sub-optimal resource allocation:

\begin{enumerate}

\item {\bf Heterogeneous resource demands of sparse and dense layers in
	RecSys.} Modern RecSys combines sparse embedding layers with dense DNN
	layers, each layer exhibiting notable differences in their compute and memory
	characteristics. Consequently, the QPS of a dense DNN layer and sparse
		embedding layer becomes uneven (i.e., one layer type typically shows much
				lower QPS than the othey layer type), rendering the low-performance layer
			to bottleneck the overall QPS
			(\sect{sect:motivation_heterogeneous}).  To maximize end-to-end
			model-wise throughput, an optimal resource allocation would have
			\emph{more compute resources provisioned just to the bottlenecked layer}.
			Unfortunately, the baseline model-wise resource allocation treats the
			entire RecSys model as one monolithic unit for allocating resources.  As
			such, it is challenging to selectively provision more resources to only a
			subset of model layer(s), failing to satisfy the unique resource
			demands of each  layer independently.

\item {\bf Sparse embedding table accesses and its skewed access distribution.} The embedding tables employed in sparse embedding layers are memory capacity limited, amounting to several tens of GBs. Access patterns to these tables generally exhibit a power-law
distribution where a small subset of (hot) table entries receive very
high access frequency while the remaining (cold) entries receive
only a small number of accesses. Because the baseline mechanism allocates resources in a 
coarse-grained, model-wise fashion,
\emph{each server replica must allocate the entire embedding tables in memory without accounting for the individual embedding's actual ``utility''}. As such, the baseline model-wise resource allocation suffers from significant memory waste and limits the total number of  server replicas that can be instantiated across the datacenter fleet, deteriorating fleet-wide QPS.

\end{enumerate}

To this end, we propose \proposed, a RecSys model serving architecture
providing resource elasticity and high memory efficiency. The unique
aspects of \proposed are twofold.
	First, our proposed system employs a
	\emph{microservice-based} inference server for high resource
	elasticity. Second, \proposed achieves high memory efficiency via our
	\emph{utility-based} resource allocation policy.

\begin{itemize}
    \item {\bf Microservice software architecture for RecSys.} \proposed employs the microservice~\cite{microservice} programming model as the core mechanism to enable fine-grained resource allocation that meets the heterogeneous resource demands of RecSys. The benefit of  microservices is that it helps partition a large monolithic application (in our case the model-wise replication of a RecSys inference server) into many fine-grained and loosely-coupled services. \proposed partitions a target RecSys model into fine-grained \emph{model shards}, each of which is implemented as a microservice and containerized for deployment. These model shards are utilized as the unit of resource allocation which allows the container orchestration system,    Kubernetes~\cite{kubernetes},
    to independently scale the number of shard replicas, providing high resource elasticity.

\item {\bf Utility-based resource allocation policy.} Our \proposed partitions the baseline monolithic RecSys model architecture into two distinct model shard types, a shard that handles dense DNN layers and sparse embedding layers. The embedding layer shard is further partitioned into hot and cold embedding shards based on the utility of embedding table entries.
Such design decision opens up unique opportunities to properly align its resource allocation with its actual utility. First, \proposed can now selectively scale-\emph{out} the number of replicas for the hot embedding shards that matches its high memory access demands. On the other hand, it also prevents \proposed from needlessly over-provisioning shard replicas servicing cold embeddings thereby minimizing resource waste. \proposed exploits these  properties to design a 
utility-based resource allocation policy  that identifies the
 optimal embedding table partitioning algorithm, which Kubernetes' autoscaling policy utilizes to replicate the appropriate number of shards to fulfill a target QPS goal.

\end{itemize}

Overall, \proposed presents a model serving architecture for RecSys
that helps customize its resource allocation best suited for its resource
demand. We demonstrate that \proposed provides an average $3.3\times$ reduction
in memory allocation size and $8.1\times$ increase in memory utility, which
reduces deployment cost by an average $1.6\times$.

\section{Background}
\label{sect:background}

\subsection{DNN-based Recommendation Models}
\label{sect:recsys_arch}

\begin{figure}[t!] \centering
\includegraphics[width=0.41\textwidth]{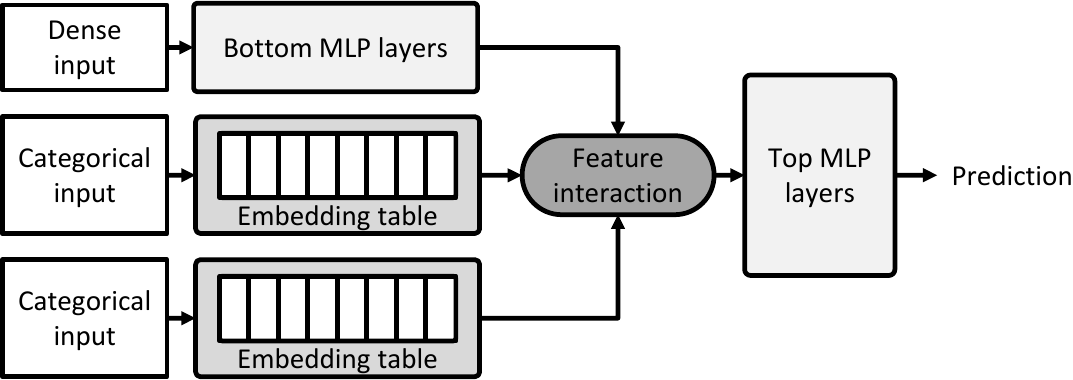}
\caption{
A modern DNN-based RecSys model architecture.
}
\vspace{-1.4em}
\label{fig:recsys_arch}
\end{figure}

Recent RecSys combines both sparse and dense features to enhance model accuracy and utilize two major components, a dense DNN layer using multi-layer perceptrons (MLP) and a sparse embedding layer (\fig{fig:recsys_arch}).
A sparse feature represents a categorical input (e.g., Ads a user has clicked in the past) and a dense feature represents a continuous input (e.g., user's age). Sparse features cannot be used directly as inputs to a dense DNN layer because they represent categorical information. As such, RecSys models use \emph{embedding tables}
to translate a given sparse feature value into a dense embedding vector. An embedding table is an array of
embedding vectors 
and
a sparse feature input (which is an index ID to the embedding table) is used to read out a particular embedding vector from
this table.
Because the number of unique items that fall under a sparse feature category can amount to several millions to billions (e.g., number of product items sold in Amazon), an embedding table can be sized at several tens of GBs. In general, multiple embeddings are \emph{gathered} from a given embedding table which are subsequently \emph{pooled} into a single embedding vector using reduction operations like element-wise additions. 

There are two distinguishing aspects of embedding layers vs. dense DNN layers.
First, the compute intensity of embedding gather and pooling operations are
extremely low, exhibiting \emph{memory bandwidth limited} behavior, especially
for embedding layers with large pooling values (number of embeddings to gather
		from a table). Second, because a modern RecSys model employs
multiple embedding tables, deploying a RecSys model causes \emph{high memory
	capacity overheads}.

\subsection{Model Serving Architectures}

\begin{figure}[t!] 
\vspace{-0.7em}
\centering
\subfloat[]{\includegraphics[width=0.23\textwidth]{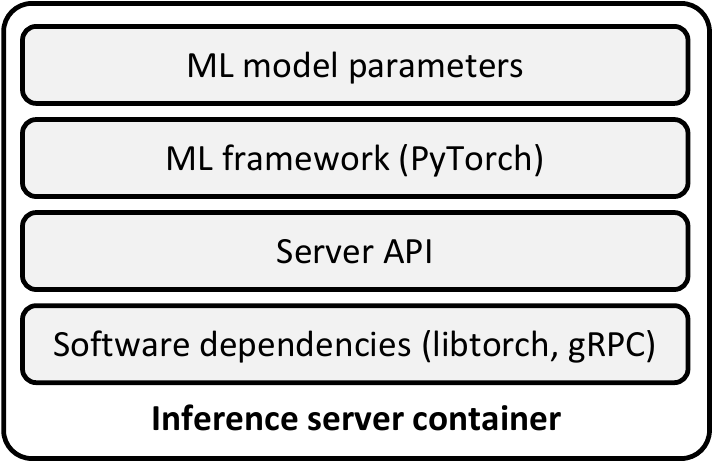}\hspace{0.5em} }
\subfloat[]{\includegraphics[width=0.23\textwidth]{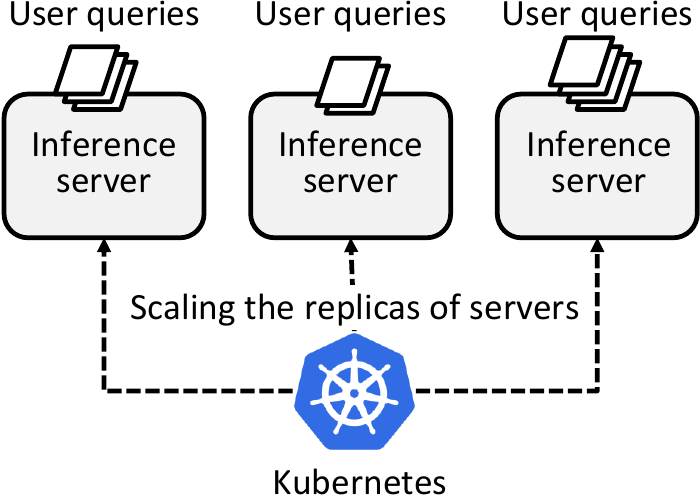}}
\caption{
(a) A containerized ML inference server using model-wise resource allocation, and (b) using Kubernetes to scale out multiple server replicas across the datacenter to meet a target QPS goal.
}
\vspace{-1.4em}
\label{fig:model_serving_arch}
\end{figure}

{\bf ML inference server design.}
Current ML inference servers utilize 
containers~\cite{container} for their deployment because of its portability and scalability (e.g., TensorFlow Serving, TorchServe~\cite{tf_serving,torch_serve}). 
A containerized ML inference server is packaged as a Docker image which contains all the essential ML software packages, the necessary system environment settings, and importantly the \emph{ML model} to deploy.
Because a ML model is treated as one \emph{monolithic} application to be serviced and containers are the smallest unit of resource allocation and deployment, any given replica of the containerized inference server must contain a copy of the \emph{entire} ML model parameters (\fig{fig:model_serving_arch}(a)). 
This paper refers to such baseline resource allocation mechanism as \emph{model-wise} resource allocation.

{\bf Deploying inference servers at scale.} The de facto standard in container
orchestration is Kubernetes~\cite{kubernetes} which helps automate the
deployment, scaling, and resource management of the containerized inference
servers at scale (\fig{fig:model_serving_arch}(b)). Kubernetes cluster
scheduler enables system designers to hand over the responsibility of resource
management by defining deployment policies. When deploying an application,
			 programmers can simply set the desired policy, and  Kubernetes
			 transparently manages the allocation of resources based on the specified
			 policy. One important, automated resource management feature provided
			 with Kubernetes is the Horizontal Pod Autoscaling (HPA)~\cite{hpa}. HPA
			 automatically adjusts the number of container replicas of inference
			 servers to satisfy a target service throughput metric (e.g., QPS) and
			 guarantee high-quality service experience to the end users.

When a new inference server is instantiated by Kubernetes, it provisions all the resources needed for that container. 
As such, an initialized inference server uploads all of its ML model parameters in memory. Such model-wise allocation of resources enables each replica of the inference server to independently service user queries.

{\bf System architectures for RecSys inference server.}
As noted in \sect{sect:recsys_arch}, the size of a single embedding table can be up to several tens of GBs. Therefore, \emph{each inference server must be provisioned with a large enough memory to store these embedding tables}. Because high-bandwidth memory employed in GPUs are not large enough to store the entire embedding tables, 
modern RecSys inference servers employ
CPU-only~\cite{deeprecsys,gupta2020architectural,hercules,lui2021understanding,facebook_hpca2018,park2018deep,jain2023optimizing}
or hybrid CPU-GPU
systems~\cite{recpipe,deeprecsys,aibox,tensorcasting,wide_and_deep,acun2021understanding,tensordimm,nvidia_cpu_gpu,merlin_hugectr}.
Both CPU-only and CPU-GPU systems share a common property
where the \emph{memory-hungry embedding tables are stored in
capacity-optimized CPU memory}. 

Therefore, unlike compute-intensive DNN layers which get executed by the GPU in
	a CPU-GPU system (i.e., CPU-only executes DNNs using the CPU), the 
	embedding layers are executed by the CPU in both CPU-only and CPU-GPU.

Hence, effectively utilizing CPU memory with
maximum efficiency becomes vital to optimize cost. This is because the total
CPU memory size determines \emph{how many} RecSys inference
servers can be deployed across the datacenter, which heavily impacts the
fleet-wide QPS. 
In this work, we study the merits of \proposed's resource allocation 
	by using \emph{both} CPU-only
and CPU-GPU based RecSys inference servers, demonstrating its wide
applicability.

\subsection{Microservices}

Microservices~\cite{microservice} break apart complex \emph{monolithic}
applications, whose functionality is implemented as a single service, into many
\emph{fine-grained} and \emph{loosely-coupled} microservices. Each microservice
is designed to serve a small subset of the original  application's
functionality,  communicating with other microservices  using Remote Procedure
Calls (RPC) or a RESTful
API~\cite{kaur2010evaluation,feng2009rest,srinivasan1995rpc,dmitry2014micro}.
A key advantage of microservices
is its elasticity. Specifically, because the granularity in which resource
allocation and scheduling are done is in individual microservices, it
facilitates deploying, scaling, and updating individual microservices
independently, improving the elasticity of resource allocation and its
scheduling.

\subsection{Related Work}
\label{sect:related}

\sethlcolor{lightgray}
\sethlcolor{yellow}
With the growing interest in
RecSys, there has been a large body of prior work exploring hardware/software
optimizations for RecSys which we summarize below.

{\bf Memory bandwidth bottleneck of RecSys.} As discussed in
\sect{sect:recsys_arch}, the embedding vector gathers and pooling operations in
RecSys incur significant memory bandwidth demands causing a  bottleneck.
Several prior work proposed near-/in-memory
processing~\cite{tensordimm,recnmp,tensorcasting,asgari2021fafnir,trim},
in-storage processing~\cite{recssd,sun2022rm} to 
alleviate embedding layer's memory bandwidth demands.
Similar to our work, there are also studies that observes and utilizes the skewed embedding
	table access pattern in RecSys for system-level optimizations.
For instance, several prior studies utilize the skewed embedding access
patterns to explore the efficacy of caching 
 to reduce overall memory bandwidth demands~\cite{
	recnmp,scratchpipe,kal2021space,merci,balasubramanian2021cdlrm,
	eisenman2019bandana, kurniawan2023evstore, xie2022fleche, yin2021tt,
	miao2021het}.
	Others leverage a heterogeneous memory
hierarchy \mbox{~\cite{recshard, ardestani2022supporting,adnan2021accelerating,song2023ugache}} to
effectively lower the average latency to access slow memory. Overall, these prior  
work alleviates the embedding layer's memory bandwidth requirements by
exploiting the skewed access patterns. \proposed, on the other hand, utilizes 
embedding table's unique access pattern to develop a cost-efficient, elastic resource management
system.

{\bf Memory capacity bottleneck of RecSys.}  Kwon et al.~\cite{tensordimm,tensorcasting} proposed a disaggregated
memory architecture to store large embedding tables in a remote memory node and
Gouk et al.~\cite{gouk2022direct}  explored the viability of CXL-based memory
pooling for storing embedding tables. Lui et al.~\cite{lui2021understanding} {explores the efficacy of distributed inference for RecSys where a model is partitioned
	and distributed 
across multiple machines for deployment. Such design point helps address
RecSys embedding layer's memory capacity demands by storing
different embedding tables in remote CPU nodes, collecting the pooled embedding
vectors using RPC calls, which is similar to} \proposed's {microservice based
design.} Mudigere et al.~\cite{mudigere2022software} explores
various model-parallel training schemes targeting embedding tables, discussing
different table partitioning plans (e.g., column-wise, row-wise, table-wise
partitioning) to address the memory capacity demands of RecSys training.
All of these prior work strictly focus on evaluating the efficacy of their
solution under a \emph{single} inference and/or training server setting without
consideration of its deployment at scale nor its resource allocation
efficiencies. More importantly, the RecSys model architecture is implemented as
one monolithic application, unlike \proposed where we focus on partitioning its
implementation into fine-grained model shards using a microservice architecture {to enable elastic resource scaling}.
DisaggRec~\cite{ke2022disaggrec} proposes a disaggregated memory
{	system augmented with near-memory processing architectures for RecSys inference, 
 which is designed to cost-effectively manage
large embedding tables. DisaggRec addresses resource
inefficiencies that arise from the varying computational and memory
requirements of dense and sparse layers. 
Similar to} \cite{lui2021understanding}, DisaggRec
{employs a distributed inference approach for RecSys and aims to tackle the resource underutilization
issue by distributing the allocation of dense layers and sparse embedding layers across compute-nodes and memory-nodes, respectively,
which helps improve machine utilization.
Unlike} \proposed's {dynamic, elastic resource management system, however, DisaggRec 
 allocates a fixed amount of resources to each layer, and determining
the optimal distribution of resources requires an exhaustive search. Overall, the key contribution of }
\proposed is orthogonal to these related work.

{\bf Runtime system and scheduling for RecSys.} While not necessarily employing microservices, there are several prior work suggesting RecSys optimized runtime systems or scheduling policies. 
Prior work \cite{hercules,deeprecsys,gupta2020architectural} suggests multi-tenant scheduling to improve the throughput of RecSys inference servers. JiZhi~\cite{jizhi} optimizes the  serving cost of RecSys by batching the model serving pipeline. MP-Rec~\cite{mprec} dynamically selects the optimal hardware platform within a heterogeneous RecSys inference server containing GPUs, TPUs, and IPUs, for better performance. In general, the key contributions of \proposed is orthogonal to these prior work.

\section{Motivation}\label{sect:motivation}

A critical limitation of model-wise allocation is that it is
difficult to flexibly allocate the appropriate amount of resources to
individual layers that match their utility and need, leading to
resource waste. This section describes the two key factors behind baseline
mechanism's sub-optimal performance.

\sethlcolor{pink}
\begin{figure}[t!] \centering
\vspace{-0.7em}
\subfloat[]{\includegraphics[width=0.27\textwidth]{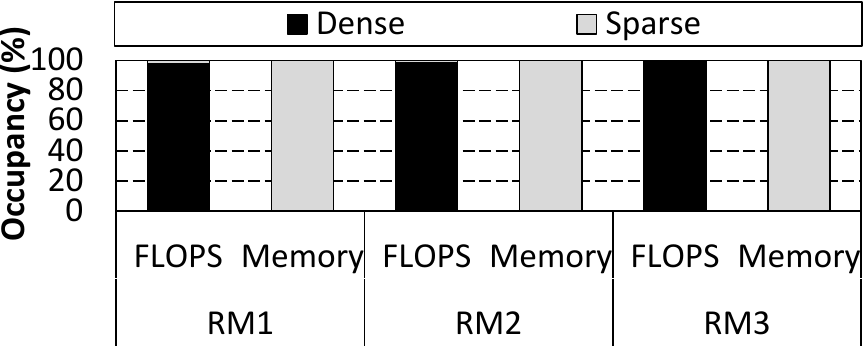}\hspace{0.1em} }
\subfloat[]{\includegraphics[width=0.185\textwidth]{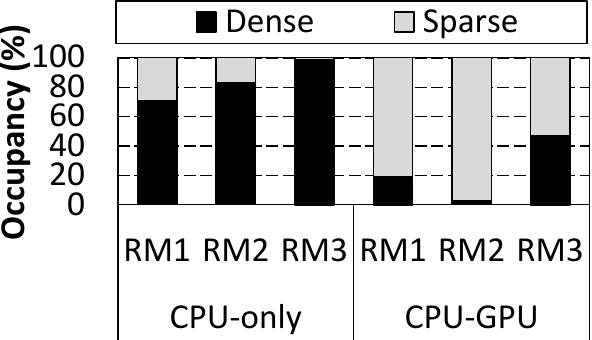}}
\caption{
The fraction of (a) FLOPs, memory consumption and (b) end-to-end inference latency (over CPU-only and CPU-GPU systems) the sparse embedding and dense DNN layers account for when evaluated over the three models studied in this paper (RM1, RM2, and RM3). FLOPs and memory consumption are architecture-independent, so its values are identical over CPU-only and CPU-GPU systems. \sect{sect:methodology} details our methodology.
FLOPS percentage of sparse embedding layers in (a) are 2\%, 1\%, and 0.1\% for
	RM1, RM2, and RM3, respectively. Memory consumption percentage of dense
	DNN layers in (a) are 0.02\%, 0.02\%, and 0.4\% for RM1, RM2, and RM3, respectively.
}
\vspace{-1.4em}
\label{fig:motivation_dense_vs_sparse}
\end{figure}

\subsection{Heterogeneous Resource Demands of RecSys}
\label{sect:motivation_heterogeneous}

The sparse embedding layer and dense DNN layer exhibit notable differences in
their compute and memory characteristics, including compute intensity (FLOPs),
			memory footprint, and memory access pattern.  Compared to large
			embedding tables, MLP's model size is only in the range of several
			MBs, yet their compute intensity is much higher than  embedding gather
			and pooling operations. For instance, in case of RM1
			(\tab{tab:benchmark}), dense DNNs account for $98\%$ of FLOPs and
			$67\%$/$19\%$ of CPU-only/CPU-GPU's end-to-end inference time, yet their model size only
			accounts for $0.02\%$ of overall memory consumption
			(\fig{fig:motivation_dense_vs_sparse}).  From a memory access pattern's
			perspective, servicing a single query requires the entire MLP parameters
			to be accessed, exhibiting $100\%$ utility of the model parameters
			allocated in memory.  In contrast, embedding layers, due to its sparse
			table access patterns, exhibit extremely low memory utility as it only
			touches $0.001\%$ (with a pooling factor of 100 per table) of the
			embedding tables per inference. This means that, on average, $99.999\%$
			of the parameters allocated in memory are of waste whenever a query
			is serviced.

  \begin{figure}[t!] \centering
\subfloat[]{\includegraphics[width=0.2\textwidth]{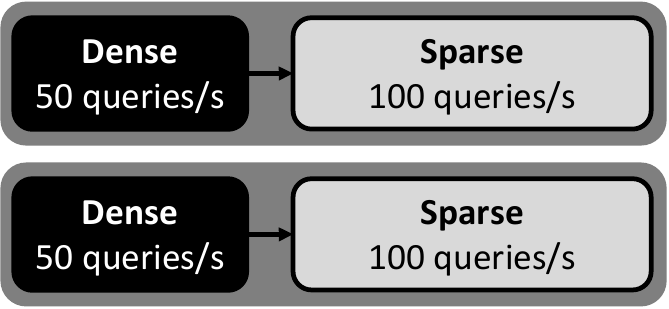}\hspace{0.5em} }
\subfloat[]{\includegraphics[width=0.2\textwidth]{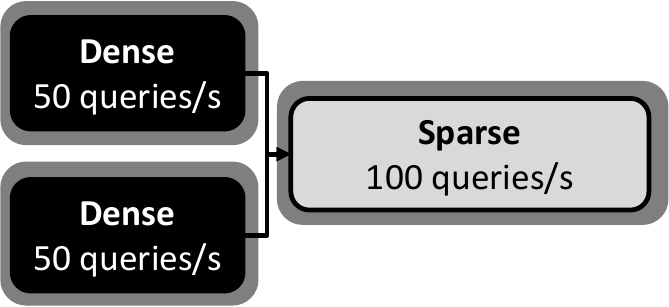}}
\caption{
An example RecSys where the dense DNN layer exhibits half the QPS than the sparse embedding layer. (a) How the baseline model-wise resource allocation would replicate two servers to reach $100$ queries/sec and (b) how our proposed \proposed would reach such QPS goal using fine-grained, per-layer resource allocation.
}
\vspace{-0.7em}
\label{fig:motivation_challenges_of_integrated_resource_allocation}
\end{figure}

\sethlcolor{green}
\begin{figure}[t!] \centering
\vspace{-0.7em}
\subfloat[CPU-only]{\includegraphics[width=0.22\textwidth]{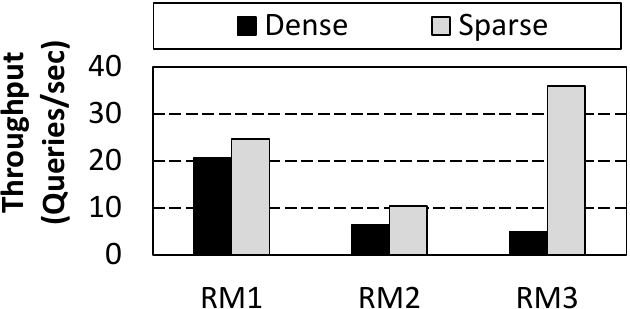}\hspace{0.2em} }
\subfloat[CPU-GPU]{\includegraphics[width=0.22\textwidth]{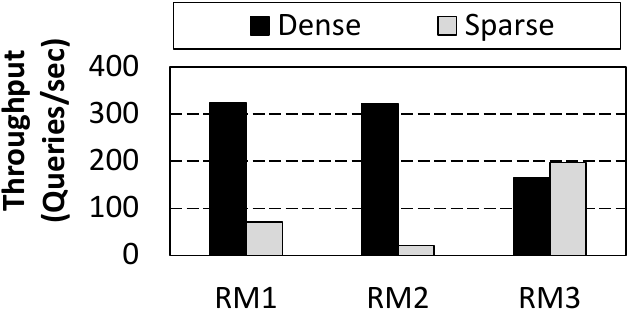}}

\caption{
Service throughput (QPS) of dense DNN and sparse embedding layers over (a) CPU-only
		and (b) CPU-GPU system when
		separately measured over the three RecSys models used in our evaluation
		(see \tab{tab:benchmark}). As shown, due to the heterogeneous resource demands of RecSys, a significant
			QPS mismatch exists between sparse and dense layers, for both CPU-only and CPU-GPU system.
}
\vspace{-1.4em}
\label{fig:throughput_sparse_dense}
\end{figure}

The mismatch in compute intensity and memory utilization, leading to
significant resource waste, cannot be addressed under the 
baseline model-wise resource allocation. Consider the example in
\fig{fig:motivation_challenges_of_integrated_resource_allocation} which assumes
that a dense DNN and sparse embedding layer can each service $50$ queries/sec
and $100$ queries/sec, respectively. Such mismatch in QPS is common in RecSys
due to its heterogeneous model architecture (\fig{fig:throughput_sparse_dense}).  
As the
DNN layer exhibits half the QPS of the embedding layer in
\fig{fig:motivation_challenges_of_integrated_resource_allocation}, the
end-to-end model-wise throughput will be bounded at $50$ queries/sec. To
increase the system-wide throughput to $100$ queries/sec, the baseline
model-wise allocation would require two replicas of the inference server to be
instantiated. From a memory efficiency perspective, such model-wise replication
is a significant waste as the  entire embedding tables are needlessly
duplicated  without contributing much to improving QPS
(\fig{fig:motivation_challenges_of_integrated_resource_allocation}(a)). A more
desirable solution would be to double the  allocated resources \emph{only} to
the dense DNN and improve its aggregate QPS to $100$ queries/sec and
resolve it from being a bottleneck
(\fig{fig:motivation_challenges_of_integrated_resource_allocation}(b)).
Unfortunately, fine-tuning the resource allocation separately on a per-layer
basis is impossible with the baseline mechanism as each inference server is
containerized as one monolithic application, forcing Kubernetes to replicate
the \emph{entire} model parameters whenever a higher system-wide QPS is desired
and a new server replica is deployed.

\subsection{Skewed Access Pattern and Locality in Embeddings}

\begin{figure}[t!] 
\vspace{-0.3em}
\centering
\subfloat[]{\includegraphics[width=0.161\textwidth]{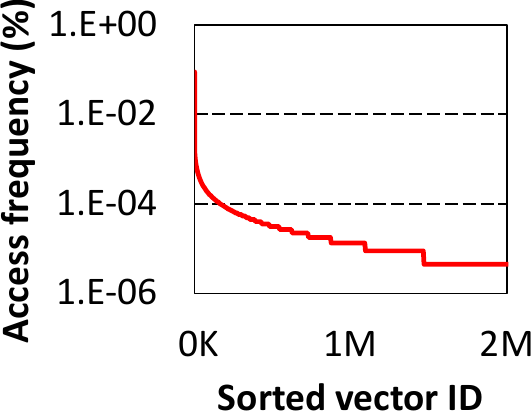}\hfill }
\subfloat[]{\includegraphics[width=0.16\textwidth]{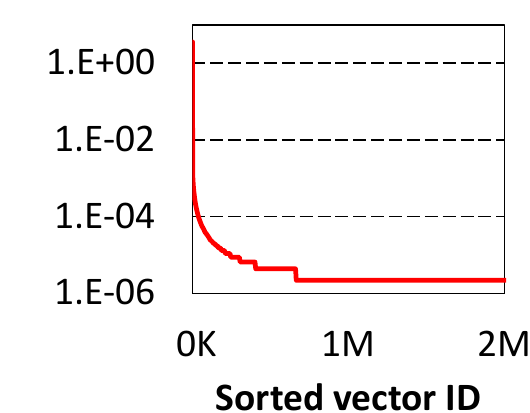}\hfill} 
\subfloat[]{\includegraphics[width=0.16\textwidth]{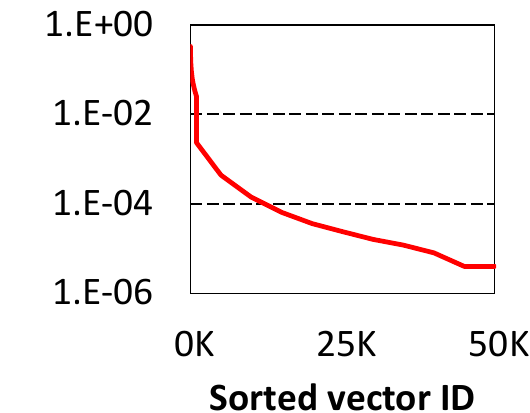}}
\caption{
Sorted access frequency of embedding vectors in real world RecSys datasets: (a) Amazon books~\cite{amazon_books}, (b) Criteo~\cite{criteo_dataset}, and (c) MovieLens~\cite{movielens}. The y-axis is plotted on a log-scale.}
\vspace{-1.4em}
\label{fig:movielens_locality}
\end{figure}

\fig{fig:movielens_locality} illustrates the access distribution of individual embedding table entries in real world RecSys datasets. As depicted, the 
table access pattern exhibits a power-law distribution where the majority
of table accesses are covered by a very small subset of the table entries (e.g., $94\%$ of accesses covered by only $10\%$ of the table entries in MovieLens).

Given such property, we can infer that an embedding layer's throughput is primarily governed by how much performance can be reaped out for embedding vector gather operations targeting those embeddings that are accessed frequently, i.e., the ``hot'' embeddings. To put it differently, in order to enhance the effective throughput of an embedding layer, it is more advantageous to selectively allocate more resources to embedding gathers targeting hot embeddings rather than cold embeddings. Unfortunately, as in our example in \fig{fig:motivation_challenges_of_integrated_resource_allocation}(a), the baseline mechanism allocates resources in a coarse-grained, model-wise manner, having 
the entire embedding tables be replicated in memory without consideration of its actual utility.

Overall, we conclude that the baseline model-wise resource allocation does not align well with the unique properties of RecSys model serving. This misalignment leads to significant waste in memory resources, which is particularly detrimental to the memory-capacity limited embedding tables.

\section{\proposed Model Serving Architecture}
\label{sect:proposed}

\subsection{Microservice-based Inference Server Design}
\label{sect:proposed_1}

\begin{figure}[t!] \centering
\includegraphics[width=0.41\textwidth]{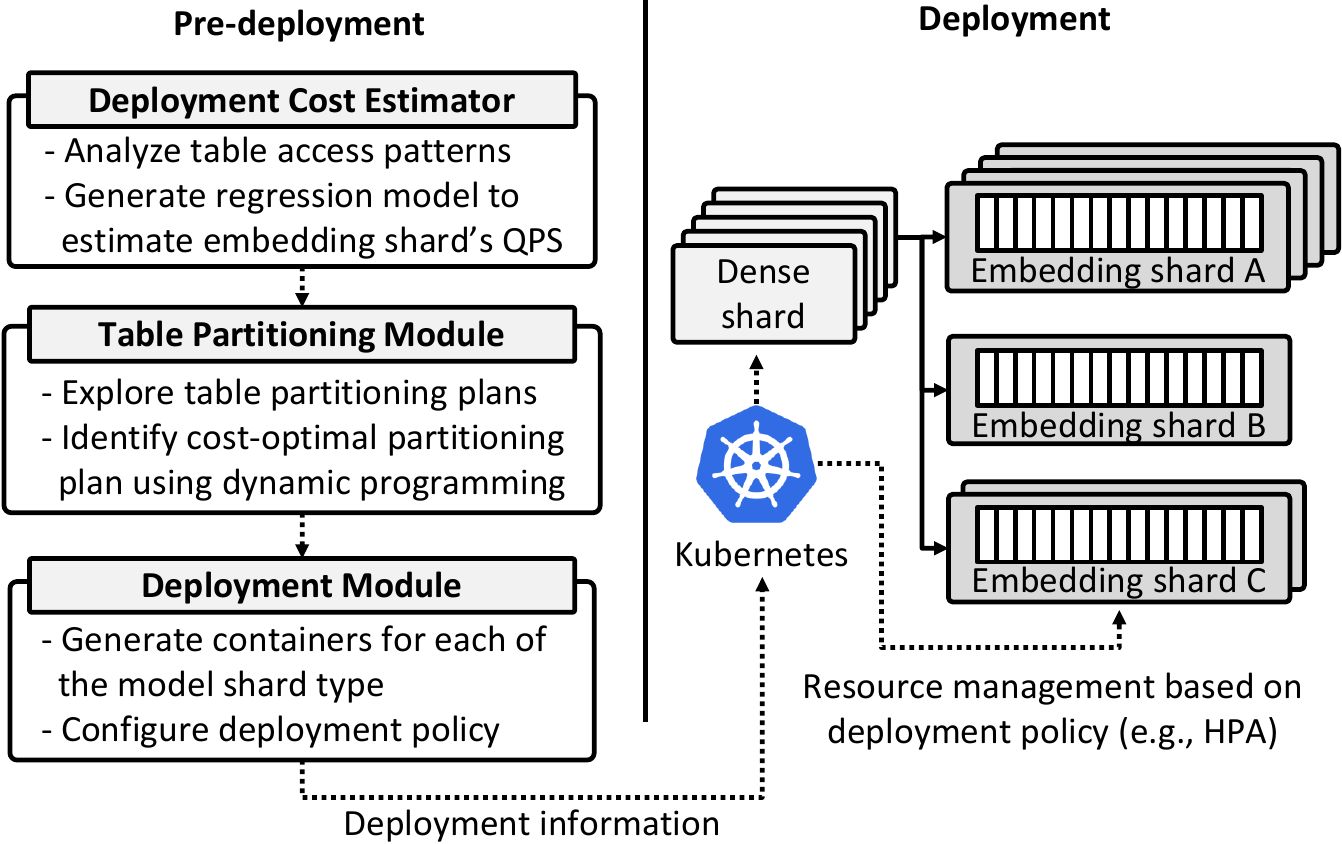}
\caption{
High-level overview of our \proposed server architecture. The example assumes that \proposed partitions the RecSys model into one dense DNN shard and three embedding shard types, each of which is containerized for deployment. To sustain a target QPS, the example assumes that Kubernetes instantiated $5$/$4$/$1$/$2$ replicas of the dense DNN and embedding shard A, B, and C, respectively.
}
\vspace{-1.3em}
\label{fig:overall_system}
\end{figure}

{\bf Server architecture overview.} \fig{fig:overall_system} provides an
overview of \proposed's model serving architecture. \proposed
employs a \emph{microservice} programming model to break down the monolithic
RecSys model serving architecture into different \emph{model shards}, each of
which is implemented as a microservice. There are two types of model
shards, a dense DNN shard and a sparse embedding shard. The dense DNN shard
services all the computations related to the bottom/top MLP and feature
interactions (\fig{fig:recsys_arch}). On the other hand, the sparse embedding
shard is responsible for gathering the requested embedding vectors stored
within that shard. An embedding table is partitioned into various sized
embedding shards based on the hotness of embeddings. The table partitioning is
done by our \emph{dynamic programming based partitioning algorithm} (detailed
		in \sect{sect:sharding_sparse}) which determines the optimal partitioning
plan that maximizes resource efficiency.

In \proposed, each model shard is containerized as a Docker image.  For CPU-only
systems, all model shards (both dense and sparse) are CPU-centric so they are
designed as containers only requiring CPU resources. As for CPU-GPU systems,
				 the containers that service sparse embedding shards are similarly
				 designed with only CPU resource requirements.  Because of dense DNN
				 layer's high compute intensity and small memory footprint
				 (\fig{fig:motivation_dense_vs_sparse}), CPU-GPU systems service
				 dense DNN shards using a GPU-centric container utilizing both CPU/GPU
				 resources.  In our design, model shard instances communicate with each
				 other using the gRPC protocol~\cite{grpc}.

{\bf Life of an inference query.} When a user query arrives to the inference
server, the input data is  routed to the dense DNN shard, which splits it into
two parts: the sparse input and the dense input. The dense DNN shard then
processes the bottom MLP layers using the dense input while concurrently
initiating RPC calls to the sparse embedding shards to collect the required
embeddings. Sending embedding gather requests across the  sparse embedding
shards requires a \emph{bucketization} process that determines which among the
partitioned embedding shards the input query should gather embeddings from
(\sect{sect:proposed_bucketization} details the bucketization algorithm).  The
embedding shards, each storing the partitioned embedding table, gather the
embeddings requested by the dense DNN shard. Once all embeddings are
gathered and pooled, the sparse embedding shards send them back to the caller
microservice, i.e., the dense DNN shard.  Upon receiving the pooled embeddings
from the sparse embedding shards, the dense shard goes through the remaining
inference process including feature interaction, top MLP, and
finally calculating the event probability which is returned back to the user.

{\bf Scaling out inference servers using Kubernetes.} In \proposed,
the containers that service dense and sparse model shards
become the unit of resource allocation and scheduling by our
container orchestration system, Kubernetes~\cite{kubernetes}. 
					 This enables \proposed to
					 independently scale the number of model shard replicas to satisfy a
					 target QPS goal, whether it be a CPU-centric shard or a GPU-centric shard, achieving high resource elasticity. Kubernetes
					 horizontal pod autoscaling (HPA) policy defines when to scale
					 up/down the number of shard replicas under what condition. In
					 \sect{sect:proposed_kubernetes}, we detail how \proposed utilizes
					 such feature to adaptively adjust the shard replica numbers
					 according to the incoming query traffic.

  \subsection{Utility-based Resource Allocation for Embeddings}
\label{sect:sharding_sparse}

The key objective of \proposed's embedding table partitioning algorithm is to determine (1) the optimal number of embedding shards to partition the table and (2) how many embeddings to include within each embedding shard, which minimizes its deployment cost while sustaining target QPS goals.
\proposed proposes a \emph{dynamic programming} (DP) based table partitioning algorithm which is based on our \emph{profiling}-based deployment cost (i.e., memory consumption) estimation model. 
We discuss each of these components below.

\begin{figure}[t!] 
\vspace{-0.7em}
\centering
\subfloat[]{\includegraphics[width=0.15\textwidth]{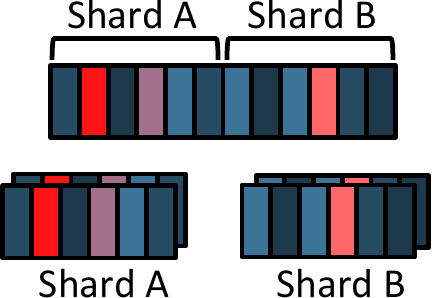}\hspace{2.5em} }
\subfloat[]{\includegraphics[width=0.153\textwidth]{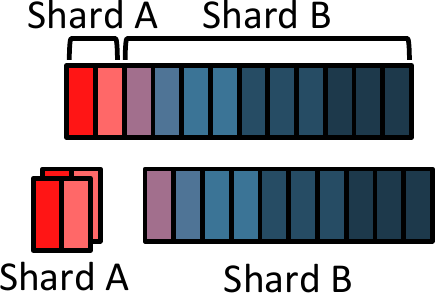}} 
\vspace{-0.5em}
\caption{
Partitioning (a) an example embedding table as-is without any preprocessing, and (b) sorting the  table first based on the hotness of embeddings and \emph{then} partitioning the table into two different shards, hot (red) vs. cold (blue) embedding shards. When it comes to partitioning an embedding table, this paper always assumes that an embedding shard includes a non-overlapping set of embeddings with consecutive index IDs.
}
\vspace{-1.3em}
\label{fig:partitioning_overview}
\end{figure}

{\bf Embedding table preprocessing.} 
As shown in \fig{fig:partitioning_overview}(a), hot embeddings are randomly dispersed across the embedding table, so partitioning the table as-is into multiple shards where each shard consists of a non-overlapping set of consecutive embeddings inevitably mixes up hot and cold embeddings altogether. Such partitioning plan reduces the effectiveness of \proposed's fine-grained resource allocation and scaling policy as cold embeddings will needlessly be duplicated whenever a new embedding shard is replicated.

To this end, \proposed first preprocesses the embedding table by sorting each
embedding's location within the table based on its access frequency. The access
frequency of an embedding can be determined by keeping a history of each
embedding's access count within a given time period, one that can easily be
implemented in production inference servers~\cite{recshard,merci}. As shown in
\fig{fig:partitioning_overview}(b), once the table is sorted, the hottest
embedding vector will be stored at the leftmost location indexed with an ID=$1$
and the coldest embedding located at the rightmost location indexed with
ID=(number of embedding vectors in table). Using the sorted embedding table, we
can now create a model shard that only includes embeddings much hotter than the
other ones, which functions as a vehicle for designing our utility-based
resource allocation policy, i.e., the ability to replicate a larger number of
model shards \emph{only} for the hot embeddings without duplicating cold
embeddings. Note that sorting the embedding table incurs a one-time latency
overhead (approximately three seconds for the largest table we evaluate) and
more importantly, such preprocessing step is off the critical path of serving
online inference queries.

\begin{algorithm}[t!]
\caption{Deployment Cost Estimation Algorithm} 
\label{algo:cost_estimation}
\begin{algorithmic}[1]
\footnotesize

\Function {Cost} {k, j} 
\State num\_replicas = REPLICAS(k, j)
\State shard\_size = CAPACITY(k, j) + min\_mem\_alloc
\State memory\_consumption = num\_replicas $\times$ shard\_size
\State \Return memory\_consumption
\EndFunction
\Statex
\Function {replicas} {k, j} 
\State{ \# \ $n_{t}$: average number of vectors to gather from the table}
\State{ \# \ $target\_traffic$: predefined constant representing user\_traffic}
\State{ \# \ $QPS(x)$: Estimated QPS of a shard that gathers $x$ embeddings,}
\Statex{ \qquad \quad which is derived by our profiling-based regression model}
\State probability = CDF(j) - CDF(k)
\State n$_{s}$ = probability $\times$ $n_{t}$ 
\Statex{\quad \ \  \# \ $n_{s}$: average number of vectors gathered from the shard}
\State estimated\_QPS = QPS(n$_{s}$)
\State num\_replicas = target\_traffic$/$estimated\_QPS
\State \Return {num\_replicas}
\EndFunction
\Statex

\Function {capacity} {k, j} 
\State \Return {$(j - k + 1) \times (size\_of\_a\_single\_embedding\_vector)$}
\EndFunction
\end{algorithmic} 
\end{algorithm}

\textbf{Deployment cost estimation.} Using the sorted embedding table,
	\proposed iterates through the evaluation space of various \emph{partitioning
		plans} (detailed in \algo{algo:dp_algorithm}) and estimates each plan's
		memory consumption to identify the optimal partitioning plan, i.e., one
		with the lowest memory consumption.  We use \algo{algo:cost_estimation} to
		explain how \proposed predicts the memory consumption of a given
		partitioning plan, which is determined by both the size of each shard
		and the number of replicas to instantiate for each shard (line 4).

The number of shard replicas to instantiate can be estimated by dividing up the
target QPS with the number of queries a specific shard is able to process
per second (line $14$). We observe that the QPS an embedding shard can sustain
is primarily determined by two key parameters: (1) the number of embeddings to
gather from that shard and (2) the size of each embedding vector which
determines the overall volume of data to fetch from memory. We employ a
\emph{profiling-based} approach to estimate these parameters as explained
below. Suppose the number of vectors to gather from the original,
non-partitioned embedding table is defined as n$_{t}$ (line 8). We first need
to predict how many embeddings will be gathered from the partitioned embedding
shard (n$_{s}$) out of the overall n$_{t}$. The value of n$_{s}$ can be
estimated by predicting what fraction of the overall table accesses
(n$_{t}$) is likely to fall under the given embedding shard. Since the
embedding vector's access frequency (one which we already used to sort and
preprocess the table) is a direct indicator of which embeddings are
most likely to be accessed to service a query, we construct a CDF (cumulative
distribution function) using the ``sorted'' embedding table's access frequency
information. Because \proposed constructs embedding shards over non-overlapping
set of embedding vectors with consecutive index IDs
(\fig{fig:partitioning_overview}(b)), an embedding shard starting from index ID
$k$ to $j$ ($k<j$) is likely to account for (CDF($j$)$-$CDF($k$)) percentage of
n$_{t}$  gathers (line 11). By multiplying this probability with n$_{t}$, we
get a reliable estimation of the value of n$_{s}$ (line 12).

\begin{figure}[t!] \centering
\includegraphics[width=0.38\textwidth]{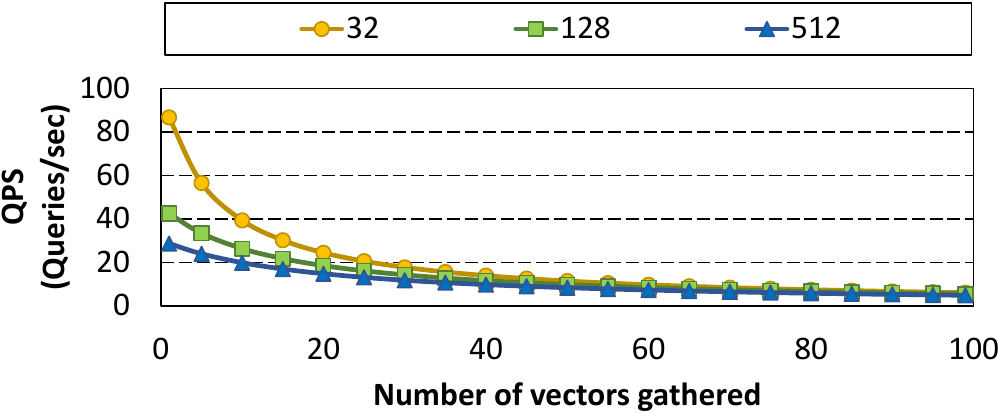}
\vspace{-0.2em}
\caption{
The changes in QPS (y-axis) as a function of the number of embedding gathers (x-axis) conducted over a 20M entry embedding table. We change the size of the embedding vector dimension (from 32 to 512 element vector) to illustrate how different data volume sizes that are fetched from memory impact the QPS, i.e., the larger the dimension size, the smaller its QPS due to higher read traffic.
}
\vspace{-1.4em}
\label{fig:num_vectors_vs_throughput}
\end{figure}

Now that we have determined the number of embeddings to gather from a shard
(n$_{s}$), we discuss how to predict the estimated QPS for that shard.  The QPS
of an embedding gather operation is determined not only by the number of
embeddings to gather from that shard but also the underlying hardware
architecture the gather operation is initiated. Given such, \proposed conducts
a one-time profiling of embedding vector gather operations, swept over various
number of vector gathers, and measures its QPS to construct a lookup table
indexed by the number of gathers (\fig{fig:num_vectors_vs_throughput}). We
utilize this profiled lookup table to generate a regression model ($QPS(x)$ in
line 10, 13) that estimates the QPS of an embedding gather operator as a
function of n$_{s}$. The estimated QPS of a shard is utilized to determine the
number of replicas required to meet a target QPS goal (line 14).
As for the target QPS goal in line 14,
it serves
	as a constant value for the dynamic programming algorithm as all the
	partitioning plans share the same QPS values. Any QPS values that make
	the number of replicas larger than 1 can be utilized for the target
	QPS. Here, we utilized 1000 for the QPS goal.
Since each
shard's memory consumption (line 3) is determined by the embedding shard size
(line 18) and other minimally required memory allocations for each container
(e.g., code, input buffers, min\_mem\_alloc in line 3), we multiply the number
of replicas (line 2) with per-shard memory consumption (line 3) to get  an
estimated  memory consumption for deploying that shard (line 4).

\begin{figure}[t!] \centering
\includegraphics[width=0.4\textwidth]{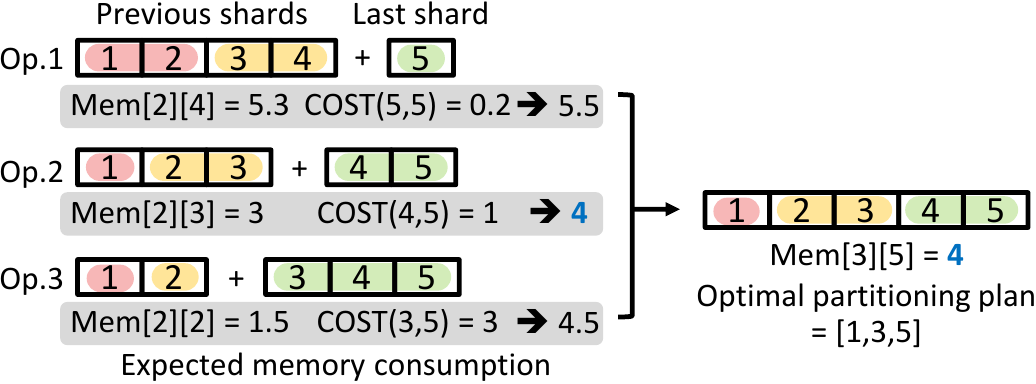}
\caption{
Example of how our DP algorithm evaluates Mem[3][5] and its optimal partitioning plan. Different shards are colored differently (red/yellow/green).
For clarity of explanation, this example assumes that COST(k,j) is defined as a simple function (unlike \algo{algo:cost_estimation}) that returns a value equal to (j - i + 1)$^{2}$/i. For instance, COST($4$,$5$) = (5 - 4 + 1)$^{2}$/4 = $1$.
}
	\vspace{-0.5em}
\label{fig:dp_example}
\end{figure}

\textbf{DP-based table partitioning algorithm. } DP is a problem-solving
technique that breaks up a complex problem into a set of sub-problems. DP
expresses the solution to the complex problem \emph{recursively} in terms of
the sub-problems and solving the recursive relation without repeatedly solving
the same sub-problem twice by \emph{memoizing} previously solved sub-problems.
We use \fig{fig:dp_example} to explain how DP
sub-problems are defined and solved for embedding table partitioning.

Consider an embedding table $E$ having N$_{max}$ embedding vectors already
sorted based on their hotness as discussed in
\fig{fig:partitioning_overview}(b). We define Mem[$num_{shards}$][$x$] as the
lowest memory cost incurred when a table $E'$ containing only the
$x$ most hot embeddings of table $E$ (i.e., $x$ $\le$  N$_{max}$, so when $x$
		equals N$_{max}$, $E'$ is equivalent to $E$) is partitioned into
$num_{shards}$ shards. In \fig{fig:dp_example}, for instance, $E$ is a
table with a total of N$_{max}$=$5$ embeddings. Also, Mem[2][3]
stores the smallest memory cost of partitioning the table $E'$ sized with $3$
most hot embeddings of $E$ (i.e., E[1,2,3]) into $num_{shards}$=$2$ shards. The
key objective of \proposed's DP algorithm is to iterate through the problem
space of Mem[$num_{shards}$][-] and identify the value of $num_{shards}$ and its
	partitioning plan that results in the least memory consumption.

\begin{algorithm}[t!]
\caption{Embedding Table Partitioning Algorithm} 
\label{algo:dp_algorithm}
\begin{algorithmic}[1]
\footnotesize
\Function {Find\_Optimal\_Partitioning\_Plan} {}

	\Statex{\hspace{0.3em} \# \ $Mem[num_{shards}][x]$: The smallest
memory cost when partitioning}
	\Statex{\hspace{0.3em}  \quad the table $E'$ including $x$ most hot embeddings to $num_{shards}$ shards}
	\Statex{\hspace{0.3em} \# \ $COST(start_{ID}, end_{ID})$: Expected memory consumption of a}
	\Statex{\hspace{0.3em} \quad  shard that contains embeddings with ID from $start_{ID}$ to $end_{ID}$}

\For {$end_{ID}$ = 1 to $N_{max}$} 
\State Mem[1][end$_{ID}$] = COST(1, end$_{ID}$) 
\EndFor

\For {$num_{shards}$ = 2 to $S_{max}$} 
\For {$end_{ID}$ = num$_{shards}$ to $N_{max}$}
\State min\_estimation $=$ float(inf)
\For {start$_{ID}$ = num$_{shards}$ to end$_{ID}$}
\State prev\_shards\_mem = Mem[num$_{shards}$-1][start$_{ID}$ - 1]
\State last\_shard\_mem = COST(start$_{ID}$, end$_{ID}$)
\State cur\_estimation = prev\_shards\_mem + last\_shard\_mem
\If {cur\_estimation $<$ min\_estimation}
\State min\_estimation = cur\_estimation
\State \emph{Memorize current partitioning points}
\EndIf
\EndFor
\State Mem[num$_{shards}$][end$_{ID}$] = min\_estimation
\EndFor
\EndFor
\State \Return partitioning points corresponding to smallest Mem value
\EndFunction
\end{algorithmic} 
\end{algorithm}

In \fig{fig:dp_example}, we illustrate the process of deriving Mem[3][5], which
represents the minimum memory cost when the table $E$ is partitioned
into three shards.  Each partitioning plan's memory consumption is the
summation of two values: (1) the estimated memory consumption of the first two
shards (red/yellow) and (2) the third (green) shard's memory consumption.
Because of the recursive relationship in DP, the optimal memory consumption for
the first two shards can be determined by referencing the memoized value of
Mem[$2$][$x$] ($x$=$4$/$3$/$2$), which represents the least memory consumed
when $E'$ is partitioned into two shards. Since the memory consumption of the
third (green) shard can be evaluated using the COST function
(\algo{algo:cost_estimation}), we arrive at the estimated memory consumption of
option $1$/$2$/$3$ as shown in \fig{fig:dp_example}. By comparing the estimated
memory consumption for each partitioning option, we can identify the
optimal solution to this problem (option $2$) that incurs the lowest memory
cost. In the given example, the first two (red/yellow) shards partitioned with
Mem[2][3] (i.e., red and yellow shards containing E[1] and E[2,3] embeddings
		respectively) and having the third (green) shard include the the remaining
two embeddings (E[4,5]) yields the least memory consumption. Thus, Mem[3][5] is
updated with a memory cost of $4$ and the \emph{partitioning points} of [1, 3,
				5] (which stores the last index ID of each shard) is separately stored
				as the corresponding, optimal partitioning plan for this example.

In \algo{algo:dp_algorithm}, we detail \proposed's table partitioning
algorithm, which is a generalization of the aforementioned example. The
initialization step of our DP algorithm partitions the table into a single
shard. Here the values of Mem[1][end$_{ID}$] represent the optimal memory
consumption when a single shard contains the end$_{ID}$ most hot embeddings
(line 2-4), one which is derived using our COST function
(\algo{algo:cost_estimation}).  The remaining Mem[num$_{shards}$][-] values are
derived by exploiting the recursive relation between the table
partitioned with ($i$ - 1) shards and the table partitioned with $i$
shards where the optimal Mem value for the ($i$ - 1) shards can always be
retrieved through the memoized solution, without re-computation (line 9).
Similar to the example in \fig{fig:dp_example}, we iterate through all possible
shard sizes for the last shard by changing start$_{ID}$, from num$_{shards}$ to
end$_{ID}$ (line 8), and evaluate its COST function (line 10) in order to
determine the overall minimum memory consumption under that partitioning plan
(line 17).  After the entire design space of Mem[-][-] is evaluated up to
maximum possible number of shards (S$_{max}$), the one with the minimum memory
cost is chosen as the final partitioning plan (i.e., the number of shards to
		partition the original table and its partitioning points).

The cost of our DP algorithm is O(S$_{max}$$\times$N$_{max}$) which can be
calculated within $18$ seconds for an embedding table with 20M entries.
Importantly, executing the DP algorithm is off the critical path of serving
online inference queries.

\subsection{Bucketization}
\label{sect:proposed_bucketization}

Since \proposed partitions an embedding table into multiple embedding shards, the index IDs used to lookup the original embedding table should be  remapped appropriately, in accordance to the partitioned embedding shards. We refer to such process as \emph{bucketization} which we explain below. 

\begin{figure}[t!] \centering
\includegraphics[width=0.41\textwidth]{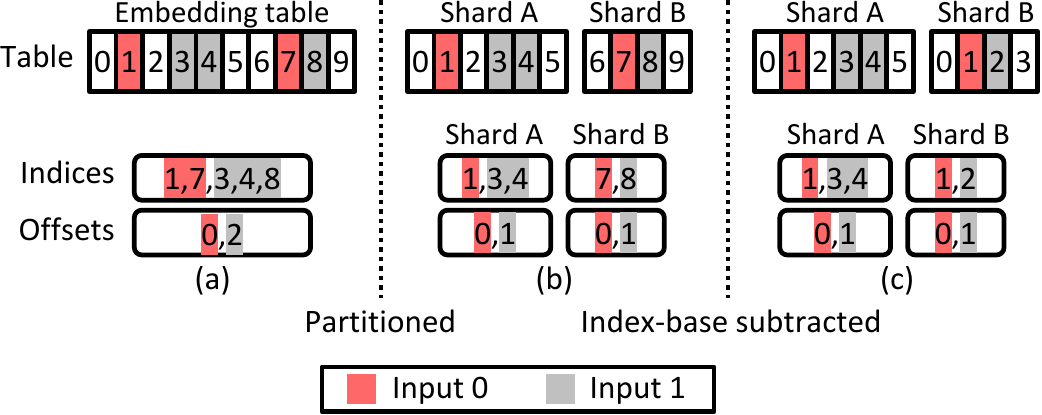}
\caption{An example bucketization process that partitions a $10$ element table into
two shards. Indices and offsets used for input 0 and 1 are highlighted in red and gray, respectively.
}
	\vspace{-0.7em}
\label{fig:bucketization_explanation}
\end{figure}

Consider the example 
in \fig{fig:bucketization_explanation} which assumes that a table with $10$ embeddings are partitioned into two shards.
To improve throughput, a single query contains multiple inputs that are batched together for concurrent processing. As such, when accessing an embedding table, two arrays are utilized, the index array and the offset array. The index array stores the list of IDs to lookup from the table, whereas the offset array is used to separate out which elements within the index array should different inputs within the query utilize. In \fig{fig:bucketization_explanation}(a), for instance, the first element in the offset array (value $0$, red) indicates that  input 0 requires index IDs starting from offset $0$ of the index array, whereas input 1 should utilize IDs starting from offset $2$ (gray) of the index array. Since these two arrays can no longer be used as-is to access the partitioned embedding shards, our proposed algorithm bucketizes the original input into two partitions as follows. First, it iterates through the original index array (and offset array) and determines which embedding shard each embedding should be gathered from, generating the intermediate index and offset arrays  as shown in \fig{fig:bucketization_explanation}(b). The  values stored in shard B's index array is then subtracted by $6$ (i.e., the size of the first shard A) so that the index IDs used to lookup the sharded table can start from a base value of $0$. The bucketization algorithm is simple to implement and highly parallelizable. We omit the pseudo-code that summarizes its implementation for brevity.

\subsection{Deploying \proposed at Scale using Kubernetes}
\label{sect:proposed_kubernetes}

Since each model shard is encapsulated as a container, Kubernetes
can independently scale each shard replicas as dictated by the HPA
policy. We employ a throughput-centric metric as the HPA target for sparse
shards while a latency-centric metricis is used for dense shards' HPA target.

		For sparse shards, we utilize shard's maximum QPS as
		the autoscaling target. Specifically, \proposed measures the
		maximum QPS each sparse shard can sustain (QPS$_{max}$),
		stress-testing each one of them by gradually increasing input
		query traffic intensity and monitoring at which point the tail
		latency increases rapidly. \proposed then configures the HPA
		policy to have each sparse shard's respective QPS$_{max}$ value
		be set as the threshold to trigger each sparse microservice to
		replicate an additional shard instance. For dense
		shards, we define a latency threshold where the auto-scaling
		HPA target is set to $65\%$ of the SLA, ensuring that service
		latency remains within acceptable bounds and does not lead to
		SLA violations. Overall, \proposed can adaptively
		adjust the replicas of each shard type that satisfies the
		demands of incoming query traffic while also achieving
		high memory efficiency.

\section{Evaluation Methodology}
\label{sect:methodology}

\subsection{Hardware Architecture}
\label{sect:methodology_hw}

{\bf CPU-only inference server.} In \sect{sect:evaluation}, we first
evaluate \proposed over CPU-only systems using a multi-node CPU cluster consisting
of one master node and eleven compute nodes. Each compute node's configuration
is in line with those employed in production CPU-only RecSys inference
servers~\cite{gupta2020architectural}.  Specifically, each CPU node is
equipped with a dual-socket Intel Xeon Gold 6242 Skylake CPU containing $32$
logical cores and $192$ GB of DRAM per socket, each socket providing $128$
GB/sec of memory bandwidth. The compute nodes communicate over a 10 Gbps
network.

{\bf CPU-GPU inference server.} We also evaluate \proposed's applicability over
CPU-GPU systems using Google Kubernetes Engine (GKE) in Google
Cloud~\cite{gke}. In this setup, we utilize a GKE cluster that contains twenty
hybrid CPU-GPU compute nodes (\texttt{n1-standard-32} node~\cite{google_platform_machine_type}
		containing 32 CPU logical cores and 120 GB of DRAM, and
		connected to an NVIDIA Tesla T4 GPU~\cite{t4} over PCIe).  The compute
		nodes communicate over a 32 Gbps network.

\subsection{Software Architecture} 

Each model shard communicates with one another using C++ gRPC protocol. RecSys
models are designed using PyTorch's libtorch (v1.12) and the DLRM GitHub
repository \cite{dlrm_github}. Resource management is handled by Kubernetes
(v1.26), which is responsible for scaling in/out each model shard replicas
according to input query traffic. Load balancing is managed using Linkerd
(v2.12), routing the input queries to the shard replicas as appropriate. We
also use a Prometheus metrics server~\cite{prometheus} to collect various
custom statistics, e.g., CPU usage, memory consumption, tail latency, and QPS.

\subsection{Workloads}

					To better illustrate \proposed's effectiveness
					on model serving, we use both microbenchmarks
					(\tab{tab:microbenchmark}) and state-of-the-art RecSys model
					configurations (\tab{tab:benchmark}) used in prior work. 

{\bf Microbenchmarks.} 
We construct several microbenchmarks using DLRM's RM1 (\tab{tab:benchmark}) as our default model configuration. The microbenchmarks are designed to better cover the large evaluation space by changing some of its key model parameters in terms of (1) the dense MLP layer size, (2) embedding table's locality, (3) number of tables, and (4) the number of shards to partition a table (\tab{tab:microbenchmark}). We use these microbenchmarks to evaluate the sensitivity of \proposed across a wide range of DLRM configurations, focusing on \proposed's effectiveness in reducing memory allocation size.

{\bf State-of-the-art RecSys workloads.} We also evaluate \proposed across
multiple dimensions in detail using three representative DLRM configurations
(\tab{tab:benchmark}) used in prior
work~\cite{naumov2019deep,deeprecsys,hercules,gupta2020architectural,lui2021understanding,mudigere2022software}.
The SLA target is set to 400ms to be consistent with industry recommendations
on SLA for RecSys, which is several hundreds of
milliseconds~\cite{deeprecsys}. All experiments are collected while
ensuring that the $95$ percentile tail latency does not violate SLA.

\begin{table}[t!]
\centering
\caption{The key parameters changed in our microbenchmark based evaluations in \sect{sect:eval:microbenchmark}. The default RecSys model configuration for our microbenchmark is based on DLRM RM1 (\tab{tab:benchmark}).}
\scriptsize
\begin{tabular}{|c|ccc|}
\hline
                                   & \multicolumn{3}{c|}{\textbf{Configurations}}                                                                                                                                                                                                                                \\ \hline \hline
\multirow{2}{*}{\textbf{\begin{tabular}[c]{@{}c@{}}MLP\\ layer\\ size\end{tabular}} }      & \multicolumn{1}{c|}{Light}                                                                   & \multicolumn{1}{c|}{Medium}                                                                     & Heavy                                                                      \\ \cline{2-4} 
                                   & \multicolumn{1}{c|}{\begin{tabular}[c]{@{}c@{}}Bottom: 64-32-32\\ Top: 64-32-1\end{tabular}} & \multicolumn{1}{c|}{\begin{tabular}[c]{@{}c@{}}Bottom: 256-128-32\\ Top: 256-64-1\end{tabular}} & \begin{tabular}[c]{@{}c@{}}Bottom: 512-256-32\\ Top: 512-64-1\end{tabular} \\ \hline \hline
\multirow{2}{*}{\textbf{Locality}} & \multicolumn{1}{c|}{Low}                                                                     & \multicolumn{1}{c|}{Medium}                                                                     & High                                                                       \\ \cline{2-4} 
                                   & \multicolumn{1}{c|}{P: 10\%}                                                          & \multicolumn{1}{c|}{P: 50\%}                                                             & P: 90\%                                                             \\ \hline \hline

\textbf{Table (N)}                 & \multicolumn{3}{c|}{\begin{tabular}[c]{@{}c@{}} Total number of embedding tables: $1$, $4$, $10$ , $16$\end{tabular}}                                                                                                                              \\ \hline \hline
\textbf{Shard}                     & \multicolumn{3}{c|}{Number of shards to partition the table: $1$, $2$, $4$, $8$, and $16$}                                                                                                                                 
                                                                                  \\ \hline
\end{tabular}
\label{tab:microbenchmark}
\vspace{-0.7em}
\end{table}

{\bf Query modeling.} A query consists of multiple items to be ranked for a given user, thus the size of a query determines the input batch size. 
We follow the methodology from prior work~\cite{deeprecsys} to model the query
distribution by setting the batch size as $32$. To model the effect of locality on embedding table
accesses, we introduce a locality metric $P$, which indicates the percentage
of total accesses that are captured by the top $10\%$ most frequently accessed
vectors (e.g., $P$=$94\%$ for MoveLens dataset, indicating that $94\%$ of
embedding table lookups are covered by the top $10\%$ hottest embeddings).
\tab{tab:microbenchmark} and \tab{tab:benchmark} shows the $P$ values in our
evaluated microbenchmarks and state-of-the-art RecSys workloads.

\begin{table}[t!]
\centering
\caption{State-of-the-art RecSys workload configurations.}
\scriptsize
\begin{tabular}{|c|c|c|c|}
\hline
                              & \textbf{RM1} & \textbf{RM2} & \textbf{RM3} \\ \hline
\textbf{Bottom MLP}           & 256-128-32   & 256-128-32   & 2560-512-32  \\ \hline 
\textbf{Top MLP}              & 256-64-1     & 512-128-1    & 512-128-1    \\ \hline
\textbf{Number of embeddings}    & 20M           & 20M           & 20M           \\ \hline
\textbf{Number of tables}     & 10           & 32           & 10           \\ \hline
\textbf{Embedding dimension}  & 32           & 32           & 32           \\ \hline
\textbf{Number of embedding gathers} & 128          & 128          & 32           \\ \hline
\textbf{Locality (P)}             & 90\%         & 90\%         & 90\%         \\ \hline
\end{tabular}
\vspace{-0.8em}
\label{tab:benchmark}
\end{table}

\section{Evaluation}
\label{sect:evaluation}

	This section evaluates \proposed over both CPU-only and CPU-GPU systems.
		For brevity and clarity of explanation, we focus our evaluation over CPU-only systems
		when studying our microbenchmarks in \sect{sect:eval:microbenchmark}.
		We then evaluate state-of-the-art RecSys workloads over CPU-only and CPU-GPU
		systems in \sect{sect:eval:benchmark} and \sect{sect:eval:gpu_eval}, respectively.

\subsection{Microbenchmarks}
\label{sect:eval:microbenchmark}

\begin{figure}[t!] \centering
\subfloat[]{\includegraphics[width=0.22\textwidth]{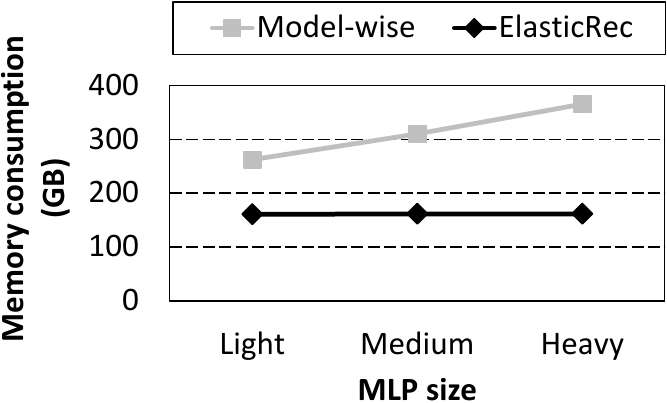}\label{fig:micro_mlp}}\hspace{0.4em}
\subfloat[]{\includegraphics[width=0.22\textwidth]{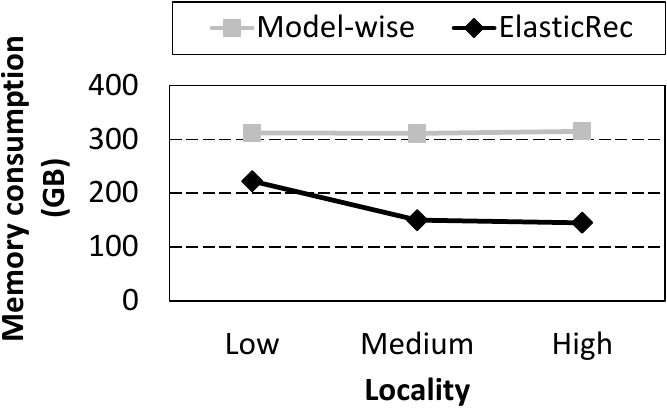}\label{fig:micro_locality}}
\vspace{0.5em}
\subfloat[]{\includegraphics[width=0.22\textwidth]{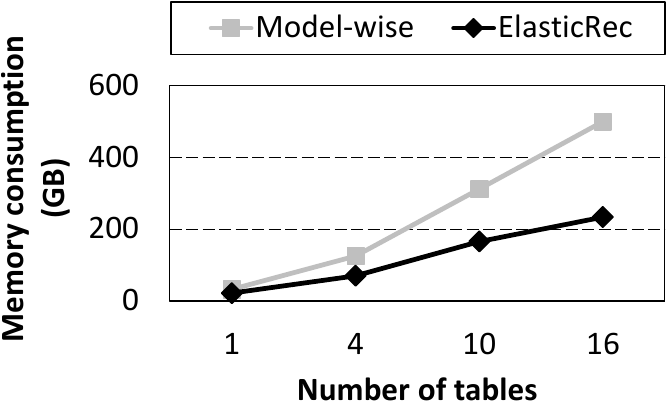}\label{fig:micro_num_tables}}\hspace{0.4em}
\subfloat[]{\includegraphics[width=0.22\textwidth]{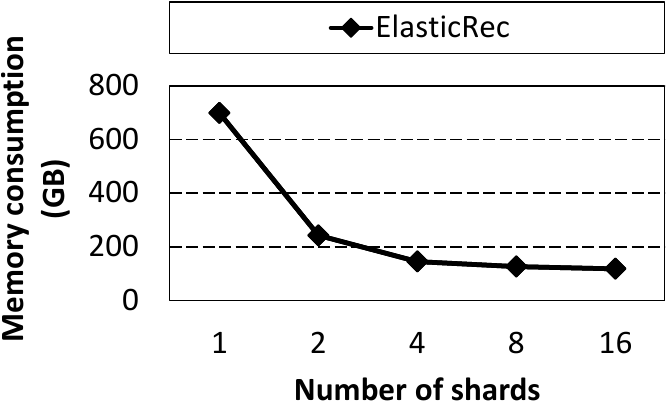}\label{fig:micro_num_shards}}
\vspace{0.0em}
\caption{ Memory consumption in our microbenchmarks, exploring the impact of (a) MLP size, (b) embedding table locality, (c) number of tables, and (d) the number of shards to partition a table.
}
	\vspace{-0.7em}
\label{fig:microbenchmark}
\end{figure}

\textbf{MLP layer size.} When the number of parameters in MLP layers is
increased (from ``Light'' to ``Heavy'' in \fig{fig:microbenchmark}(a)), the MLP
layers become more compute-intensive and experiences lower QPS. To meet the
target system-wide QPS goal,  model-wise allocation must instantiate additional
server replicas which in turn ends up duplicating the entire embedding tables.
Consequently, as the MLP layer's compute requirement increases, the overall
memory consumption under model-wise allocation also  increases rapidly. In
contrast, when the MLP size is increased to ``Heavy'', \proposed is able to
provision additional resources only to the bottlenecked MLP layers, allowing
only a modest increase in memory consumption.

\textbf{Locality in embedding tables.} 
We discussed in \sect{sect:proposed} that \proposed can allocate more resources only to those embedding shards that are accessed more frequently. 
As shown in \fig{fig:microbenchmark}(b), when the locality in table accesses is ``High'', \proposed instantiates a larger number of replicas for the hot embedding shards while spawning a relatively smaller number of cold embedding shards. Such feature helps \proposed minimize wasted memory resources allocated for servicing embedding that are not accessed frequently,  achieving $2.2\times$ memory consumption savings when locality is ``High''. The baseline model-wise allocation, on the other hand, is not able to save memory allocations at all by exploiting the table's locality, experiencing almost a constant memory consumption regardless of the level of locality.

\textbf{Total number of tables.} Recent large-scale RecSys model architectures contain a large number of sparse features, which translates into a large number of embedding tables. The microbenchmarks in \fig{fig:microbenchmark}(c) is designed to demonstrate the  scalability of \proposed's table partitioning algorithm when the number of tables is increased (the experiment assumes that all tables are sized identically, i.e., the larger the number of tables, the larger its aggregate memory consumption).
When a model contains multiple tables, \proposed applies its table partitioning algorithm separately for each individual table. For instance, if \proposed's partitioning algorithm decides that an embedding table should be partitioned into 4 shards and there exists 10 tables, a total of 40 shards (4 shards $\times$ 10 tables) will be generated, each of which will be subject for resource allocation independently by Kubernetes. Such fine-grained resource management provides \proposed with high scalability to multiple tables, showing a large performance gap against baseline model-wise allocation.

{\bf Number of shards to partition a table.} 
In \sect{sect:sharding_sparse}, we discussed how our table partitioning algorithm identifies the optimal number of shards to partition a table. To demonstrate the effectiveness of \proposed's table partitioning, \fig{fig:microbenchmark}(d) shows the overall memory consumption when we manually change the number of partitioned shards. As depicted, as the number of shards increases, the memory consumption generally decreases. Note that the memory consumption plateaus at $4$ shards, a point which \proposed's table partitioning algorithm also determines as the optimal partitioning plan to minimize memory consumption. As discussed in \sect{sect:sharding_sparse}, every container replica incurs a minimally required memory consumption (e.g., code, input buffers) to prevent containers from an out-of-memory error. As such, having an excessively large number of container replicas adds high memory overheads, leading to diminishing returns.

\subsection{State-of-the-art RecSys Workloads (CPU-only)}
\label{sect:eval:benchmark}

\begin{figure}[t!] \centering
\includegraphics[width=0.40\textwidth]{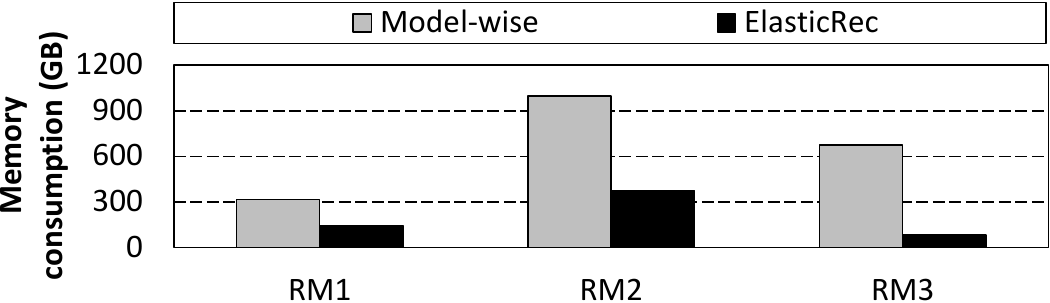}
\caption{
	CPU-only system's memory consumption over three state-of-the-art RecSys models (100 queries/sec).
}
\vspace{-0.5em}
\label{fig:dlrm_memory}
\end{figure}

\begin{figure}[t!] \centering
\includegraphics[width=0.41\textwidth]{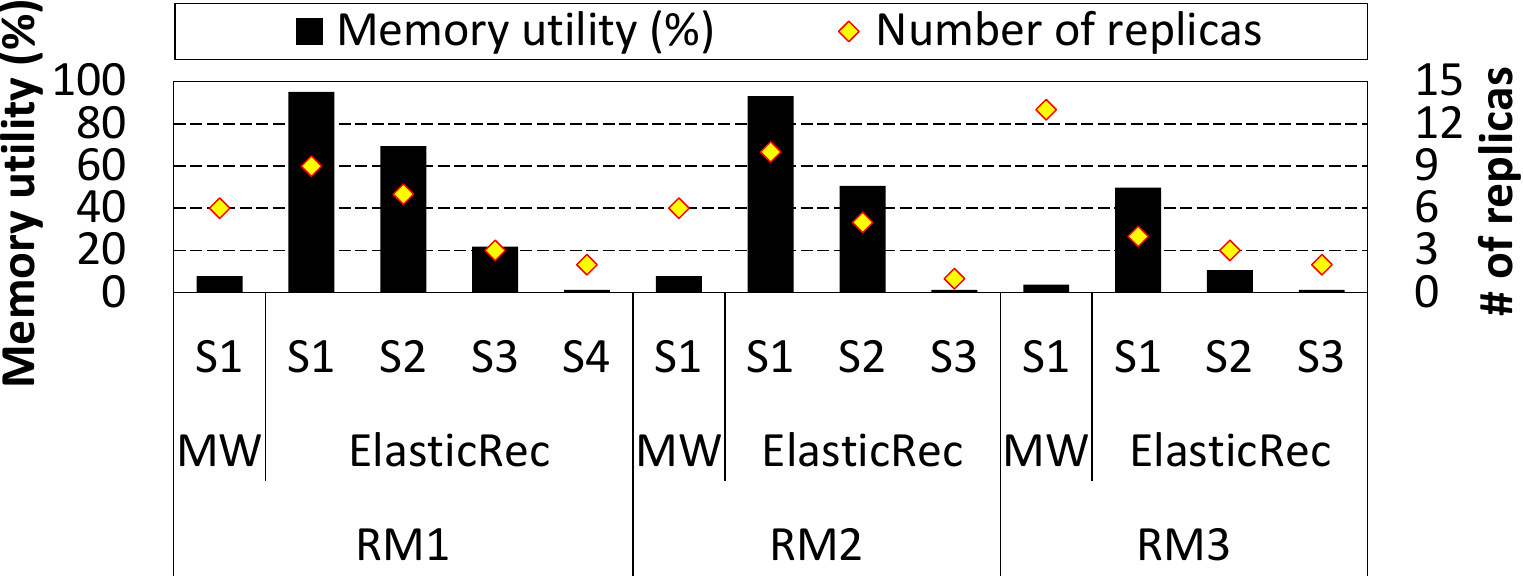}
\caption{
CPU memory utility (left axis) and 
number of shard replicas instantiated to meet target QPS (right axis) in CPU-only system. 
}
\vspace{-0.5em}
\label{fig:dlrm_utility}
\end{figure}

{\bf Memory consumption.} \fig{fig:dlrm_memory} shows the overall memory
consumption when both model-wise allocation (denoted ``MW'') and \proposed
allocate server resources to meet the same target QPS goal. For each of RM1,
				 RM2, and RM3, \proposed's partitioning algorithm decides to partition
				 the embedding tables into $4$, $3$, and $3$ shards, respectively. As
				 such, a total of $40$ shards (4 shards $\times$ 10 tables), 96 shards
				 (3 shards $\times$ 32 tables), and 30 shards (3 shards $\times$ 10
						 tables) for RM1, RM2, and RM3 (see \tab{tab:benchmark}) are
				 generated for each model's deployment, allowing Kubernetes to flexibly
				 tune the number of shard replicas in a fine-grained manner.  Overall,
				 \proposed shows substantial reduction in memory usage, achieving
				 $2.2\times$, $2.6\times$, $8.1\times$ reduction in memory consumption.
				 Note that \proposed's  memory saving is particularly significant for
				 RM3. This is because the MLP layers in RM3 are much more
				 compute-intensive than the other two models, causing the model-wise
				 allocation to replicate a larger number of inference servers as a
				 whole (detailed later in \fig{fig:dlrm_utility}) and suffer more from
				 needless duplication of cold embedding vectors.

\begin{figure}[t!] \centering
\includegraphics[width=0.40\textwidth]{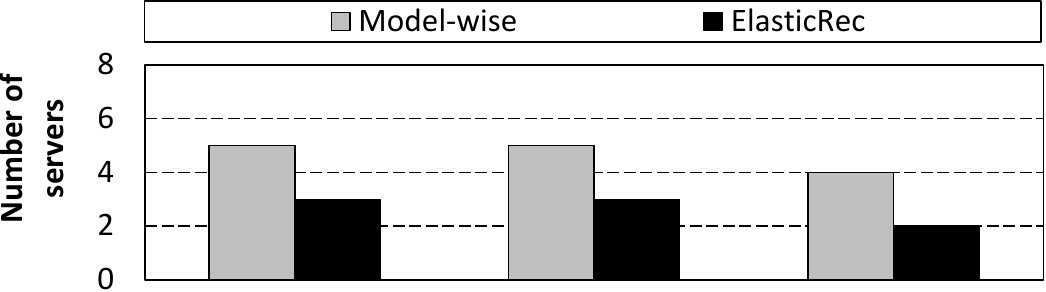}
\caption{
The number of CPU server nodes required to meet the same QPS target (100 queries/sec) in CPU-only system. 
}
\vspace{-0.6em}
\label{fig:dlrm_tco}
\end{figure}

{\bf Memory utility.} \proposed's significant memory reduction
can be attributed to intelligently allocating memory resources based on
its actual utility. We demonstrate how well memory is utilized by measuring
the percentage of embeddings that are actually accessed within a shard
while servicing the first 1,000 queries. \fig{fig:dlrm_utility} illustrates the
memory utility of each embedding shard. For brevity, we only show the utility
of the first embedding table from each workload. Each embedding shard is
denoted as S($N$), where $N$ represents the shard ID.  For \proposed, embedding
shards with a smaller ID contains hotter embeddings (e.g., embeddings in
		S1 are hotter than those in S2). Because model-wise allocation does not
partition the embedding tables, a single embedding shard exists that includes
the entire embeddings (denoted S1 under ``MW''). On average, model-wise
allocation achieves only $6\%$ of memory utility. Despite such low memory
utility, model-wise allocation must replicate the entire inference server to
meet target QPS, substantially wasting memory. Such problem becomes especially
more pronounced for the compute-intensive RM3, leading to a large number of
replicated servers and high memory consumption (as discussed in
		\fig{fig:dlrm_memory}). With our \proposed, hotter shards consistently
exhibit higher memory utility. More importantly, the number of shard replicas
is proportional to the hotness of each individual shard,
		 allowing \emph{memory resources to be preferentially allocated to those
			 shards that will actually utilize it efficiently}. Overall, \proposed
			 achieves an average $8.1\times$ higher memory utility.

{\bf Cost.} We quantify \proposed's  cost
savings by measuring the total number of CPU servers required to satisfy the
same target throughput of $100$ QPS (\fig{fig:dlrm_tco}). 
While the additional communication overheads of \proposed adds 31 ms of average latency ($8\%$ of SLA)\label{cpu_latency}, our proposal
demonstrates its efficiency by cutting down the number of deployed servers
($1.67\times$, $1.67\times$, $2.0\times$ reduction vs.
 model-wise allocation for RM1/2/3, respectively) and substantially reducing
cost by an average $1.7\times$ vs.  model-wise allocation.  These results
highlight the practical benefits and cost-efficiency of \proposed's
utility-based resource allocation.

\subsection{State-of-the-art RecSys Workloads (CPU-GPU)}
\label{sect:eval:gpu_eval}

We now demonstrate \proposed's effectiveness over CPU-GPU systems. In
\proposed, containers that service sparse embedding shards are designed with
only CPU resource requirements while compute-intensive dense DNN shards are
designed as GPU-centric containers utilizing \emph{both} GPU and CPU resources.
The baseline model-wise allocation, on the other
hand, encapsulates all CPU (sparse embedding layers) and GPU (dense DNN layers)
resources in a single container, having coarse-grained resource allocation.
Below we evaluate \proposed's effect on memory consumption/utility and cost.

\begin{figure}[t!] \centering
\includegraphics[width=0.40\textwidth]{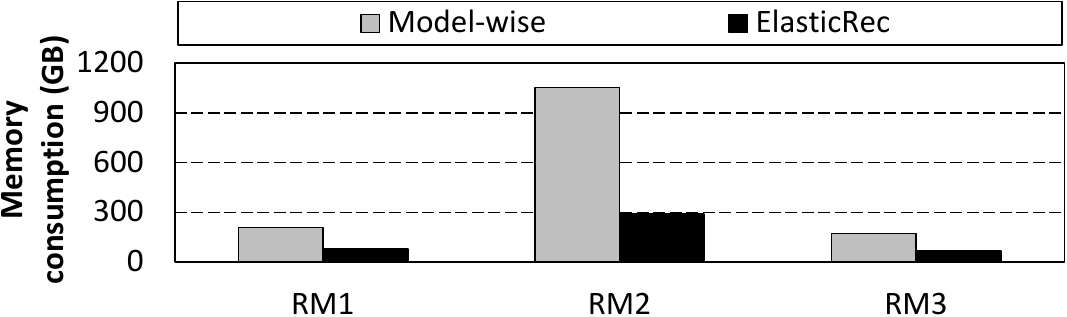}
\caption{
	CPU-GPU system's memory consumption over three state-of-the-art RecSys models (200 queries/sec).
}
\vspace{-0.5em}
\label{fig:dlrm_memory_gpu}
\end{figure}

{\bf Memory consumption.} In our CPU-GPU server, the CPU architecture
specification is different vs.  our CPU-only setting
(\sect{sect:methodology_hw}).  As such, \proposed's partitioning algorithm
decides to partition all the embedding tables into $3$ shards per table for all
three models, amounting to a total of $30$ shards, $96$ shards, and $30$ shards
for RM1, RM2, and RM3, repectively. \fig{fig:dlrm_memory_gpu} summarizes
\proposed's effect on memory consumption. It is worth pointing out that the
benefit of \proposed's memory consumption saving for RM3 ($2.6\times$
reduction) is less pronounced compared to CPU-only systems ($8.1\times$
reduction). RM3 has relatively larger MLP layers than RM1/2 so it leads to
lower QPS in a CPU-only system, necessitating a larger number of replicas to fulfill its
compute/memory demands. With CPU-GPU systems, these compute-intensive dense
DNNs are offloaded to the GPU and are executed more efficiently, requiring less
replicas. As such, the inefficiency of duplicated resource allocation is
alleviated under CPU-GPU systems which leads to a smaller gap in memory
consumption between baseline and \proposed.  Nonetheless, \proposed still shows
significant reduction in memory usage, achieving $2.7\times$,
$3.6\times$, $2.6\times$ smaller memory allocation size.

\begin{figure}[t!] \centering
\includegraphics[width=0.41\textwidth]{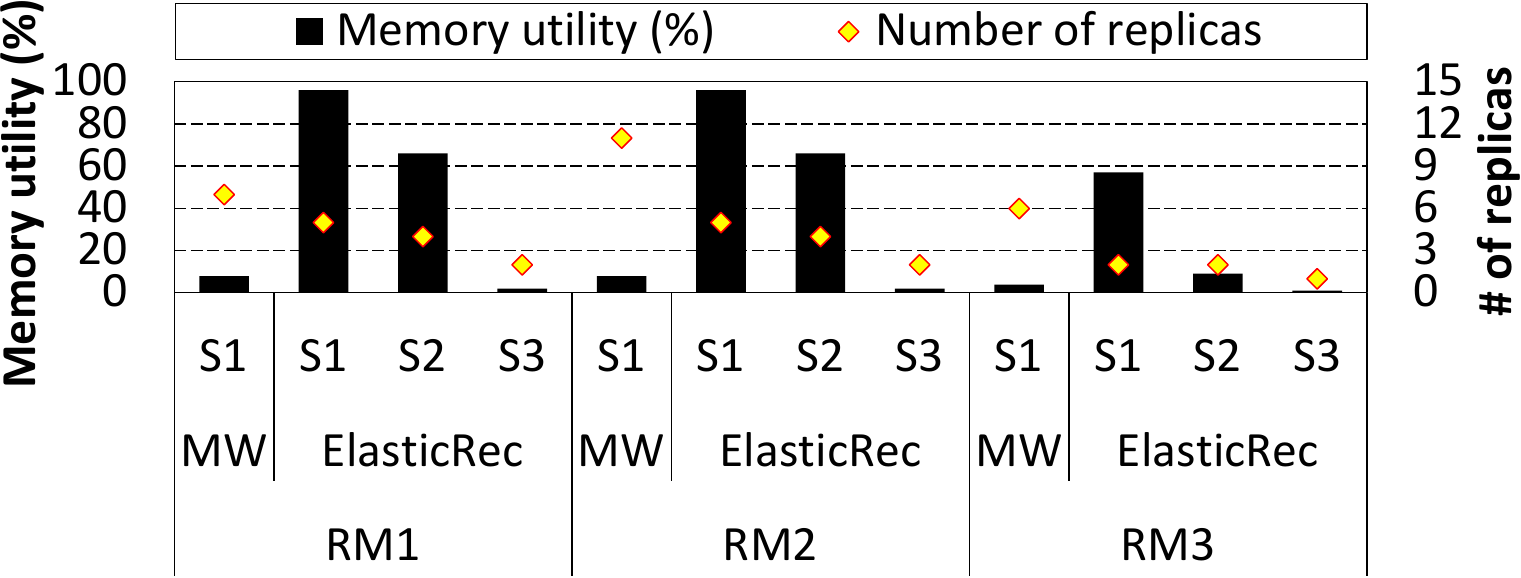}
\caption{
CPU memory utility (left axis) and 
number of shard replicas instantiated to meet target QPS (right axis) in CPU-GPU system. 
}
\vspace{-0.5em}
\label{fig:dlrm_utility_gpu}
\end{figure}

\begin{figure}[t!] \centering
\includegraphics[width=0.40\textwidth]{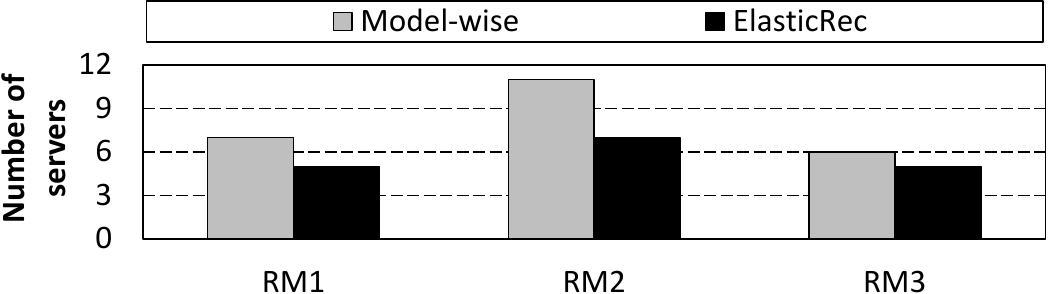}
\caption{
The number of CPU-GPU server nodes required to meet the same QPS target (200 queries/sec) in CPU-GPU system. 
}
\vspace{-0.6em}
\label{fig:dlrm_tco_gpu}
\end{figure}

{\bf Memory utility.} Similar to CPU-only systems, baseline model-wise allocation
still suffers from significant memory underutilization averaging $6\%$ memory utility
(\fig{fig:dlrm_utility_gpu}). \proposed again demonstrates its effectiveness in improving
memory utility where hotter shards consistently exhibit higher memory utility. 
Furthermore, the number of shards replicated is proportional to the hotness of
each individual shard, making sure that memory resources are allocated to those
shards that actually utilize it effeciently. On average, \proposed achieves
an average $8\times$ higher memory utilization.

{\bf Cost.}
\fig{fig:dlrm_tco_gpu} shows the number of CPU-GPU server nodes needed to
reach a target throughput of $200$ QPS. 
While the additional communication overheads of \mbox{\proposed} adds 60 ms of average latency ($15\%$ of SLA),
\proposed requires
$1.4\times$, $1.6\times$, $1.2\times$ fewer servers for RM1, RM2, and RM3, respectively,
	than baseline model-wise allocation. Overall, these results highlight the
	wide applicability of \proposed across different hardware platforms.

\subsection{Effectiveness to dynamic input query traffic}
\label{sect:eval_dynamic_traffic}

\begin{figure}[t!] \centering
\includegraphics[width=0.485\textwidth]{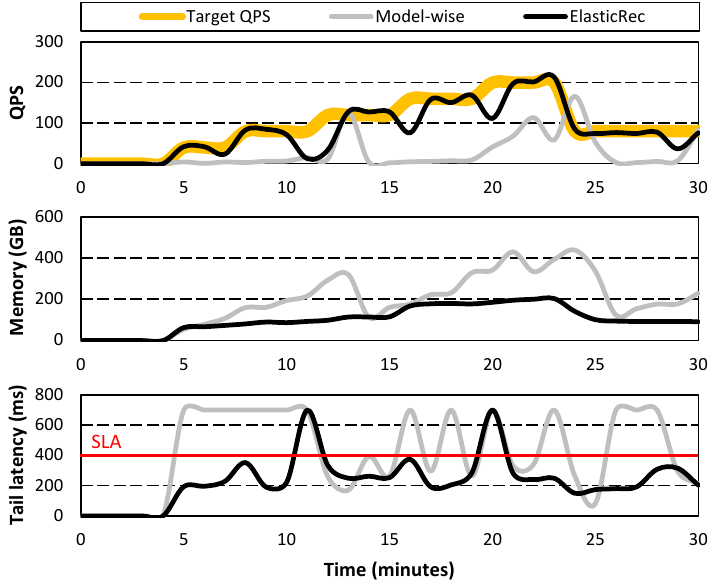}
\caption{Changes in QPS, memory consumption, and tail latency in response to the fluctuation in input traffic. The yellow line represents the target QPS in response to the input traffic. As discussed in \sect{sect:methodology}, the SLA target is set to 400ms. For brevity, we only show the results over a CPU-only system.
}
\vspace{-1em}
\label{fig:dlrm_dynamic}
\end{figure}

At-scale datacenters have a
constantly changing input query traffic, necessitating Kubernetes to adaptively
adjust the number of inference server replicas to deploy.  In
\fig{fig:dlrm_dynamic}, we demonstrate the robustness of \proposed to
dynamically changing target QPS goals when executing RM1.  We collect the resulting QPS achieved
with baseline and \proposed, its memory consumption, and tail latency. The
input query traffic is changed in a total of $5$ increments, from Time=$5$
until Time=$20$, and then decreased at Time=$24$. As the input traffic changes,
			Kubernetes scales in/out the number of replicas based on the underlying
			HPA policy. For every change in target QPS, \proposed's achieved QPS
			slightly drops in response to the traffic change and the accompanied
			change in the deployed shard replicas. However, after the shard replicas
			are appropriately provisioned, \proposed is able to quickly reach the
			target QPS goal while also meeting the tail
			latency in a stable manner. The baseline
			model-wise allocation, on the other hand, exhibits several shortcomings
			as follows.  First, the amount of memory allocated is significantly
			higher with baseline, reaching $3.1\times$ higher memory consumption than
			\proposed at its peak usage. Second, model-wise allocation responds much
			more slowly than \proposed to reach the target QPS (e.g., the QPS of
					model-wise starts to increase at around Time=$20$), exhibiting much
			more frequent spikes in tail latency that violates SLA (400ms).  These
			drawbacks arises because the granularity of resource allocation is much
			more coarse-grained under model-wise allocation, taking more time to
			initialize an inference server, load the model parameters into memory,
			and get ready to service queries. Overall, these experiments illustrates
			the \proposed's ability to effectively adjust its resource allocation to
			the dynamically fluctuating input query traffic.

\subsection{\proposed vs. GPU Embedding Caches}
\label{sect:eval_gpu_cache}

\begin{figure}[t!] \centering
\includegraphics[width=0.485\textwidth]{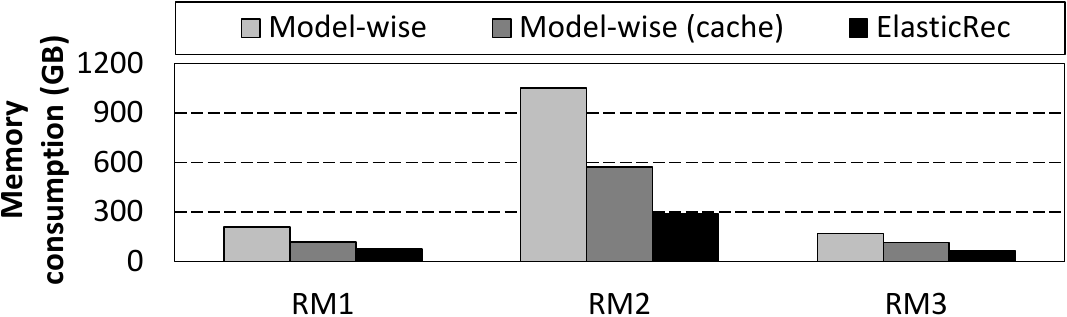}
	\caption{CPU-GPU system's memory consumption (200 queries/sec).}
\vspace{-1.4em}
\label{fig:modelwisecache}
\end{figure}

As mentioned in \mbox{\sect{sect:related}}, 
		there exists prior work that utilizes the skewed embedding table access patterns to cache hot embedding vectors inside a GPU-side embedding cache, which helps alleviate the CPU memory bandwidth pressure of embedding table lookups and increase embedding layer's throughput.
In this section, we compare \proposed's fine-grained resource management vs. baseline monolithic model-wise 
resource management augmented with a GPU-side embedding cache. In
	\mbox{\fig{fig:modelwisecache}},
the baseline model-wise allocation augmented with a GPU embedding cache is
denoted as ``model-wise (cache)''. Depending on the
size of the GPU-side embedding cache (which must be implemented inside GPU's
capacity-constrained HBM), the amount of embedding table lookup operations
captured within the GPU's embedding cache (HBM) can vary significantly (e.g., \mbox{\cite{scratchpipe}} reports that a GPU-side embedding cache that uses up to
$20\%$ of GPU's 32 GB HBM can capture $40\%$ to $90\%$ of embedding table accesses in GPU
memory). The
purpose of this study is to evaluate the implication of GPU-side embedding
cache on memory consumption savings and its overall competitiveness vs.
\proposed, so we conservatively model the baseline ``model-wise (cache)'' as
follows. Following the methodology by Kwon et al.\mbox{~\cite{scratchpipe}} we
assume that model-wise (cache) contains a large enough cache to always capture
$90\%$ of its embedding gather operations within GPU's local memory while the
remaining $10\%$ of embedding gathers are serviced from the CPU.
Compared to baseline model-wise, model-wise (cache) is able to reduce the average latency for embedding layer's execution by 47\%, leading to an increase in each shard instance's throughput
and thereby reducing the total system-wide memory consumption by 41\%.
However, the challenges of the coarse-grained model-wise resource allocation still
remains with model-wise (cache), allowing \proposed to reduce overall memory
consumption by $1.7\times$ vs. model-wise (cache).

\section{Conclusion}

We present \proposed, a RecSys model serving architecture providing
resource elasticity and high memory efficiency. \proposed overcomes the
limitations of conventional model-wise resource allocation by employing a
microservice software architecture to partition a RecSys model into
fine-grained model shards, which act as the unit of resource allocation. By
independently scaling the number of shard replicas, we demonstrated how
\proposed effectively addresses the heterogeneous resource demands of sparse
and dense layers in RecSys.

\section*{Acknowledgements}
This work was supported by the National Research Foundation of Korea (NRF) grant funded by the Korea government (MSIT) (NRF-2021R1A2C2091753). Minsoo Rhu is the corresponding author.


\bibliographystyle{IEEEtranS}
\bibliography{references}

\end{document}